\documentclass[11pt, onecolumn]{article}
\usepackage{natbib}
\usepackage{graphicx}
\usepackage{epsfig}
\usepackage{epstopdf}
\usepackage{amssymb}
\usepackage{mathrsfs}
\usepackage{url}
\usepackage{titlesec}
\usepackage{fancyhdr}

\begin{document}
\begin{center}
\Large Chemical evolution of the gas in C-type shocks in dark clouds

\vspace{0.5cm}
\large A.V. Nesterenok $^{1 *}$
\vspace{0.5cm}

\normalsize $^1$ Ioffe Physical-Technical Institute, Politekhnicheskaya St. 26, Saint~Petersburg, 194021 Russia

$^*$ e-mail: alex-n10@yandex.ru
\end{center}

\begin{abstract}
A magnetohydrodynamic model of a steady, transverse C-type shock in a dense molecular cloud is presented. A complete gas--grain chemical network is taken into account: the gas-phase chemistry, the adsorption of gas species on dust grains, various desorption mechanisms, the grain surface chemistry, the ion neutralization on dust grains, the sputtering of grain mantles. The population densities of energy levels of ions CI, CII and OI and molecules H$_2$, CO, H$_2$O are computed in parallel with the dynamical and chemical rate equations. The large velocity gradient approximation is used in the line radiative transfer calculations. The simulations consist of two steps: (i) modelling of the chemical and thermal evolution of a static molecular cloud and (ii) shock simulations. A comparison is made with the results of publicly available models of similar physical systems.

The focus of the paper is on the chemical processing of gas material and ice mantles of dust grains by the shock. Sputtering of ice mantles takes place in the shock region close to the temperature peak of the neutral gas. At high shock speeds, molecules ejected from ice mantles are effectively destroyed in hot gas, and their survival time is low -- of the order of dozens of years. After a passage of high-speed C-type shock, a zone of high abundance of atomic hydrogen appears in the cooling postshock gas that triggers formation of complex organic species such as methanol. It is shown that abundances of some complex organic molecules (COMs) in the postshock region can be much higher than in the preshock gas. These results are important for interpretation of observations of COMs in protostellar outflows.
\end{abstract}


\section{Introduction}
Around 200 different molecules have been detected in interstellar space, not counting isotopologues, as listed on the Cologne Database for Molecular Spectroscopy\footnote{\url{https://www.astro.uni-koeln.de/cdms}}. One of the motivations for studying the formation of organic molecules in interstellar medium is to trace potential chemical pathways to the formation of biologically important molecules \citep{Garrod2013}. Models based on the gas-phase chemistry alone are unable to reproduce abundances of many complex molecules observed in cold molecular clouds and their cores \citep{vanDishoeck2014}. In particular, such species as methanol as well as many more complex species are effectively formed on the surface of dust grains \citep{Watanabe2002}. The gas-phase abundance of these molecules can be enhanced by orders of magnitude due to ejection of these species from icy mantles of dust grains in shocks \citep{Lefloch2017}.

Shocks are produced in the interstellar medium in a variety of sources. (i) During a supernova event the matter thrown away forms a shock wave that propagates in the interstellar medium and interacts with molecular clouds. Studies of molecular emission from the clouds encountered by a supernova remnant can provide information about magnetic field strength, density and velocity of the gas \citep{Wardle2002}. (ii) Complex organic molecules (COMs) have been detected in outflows powered by low-mass and high-mass protostars -- it is suggested that these species are ejected during the sputtering of grain mantles in shocks \citep{Palau2017}. (iii) It has been proposed by \citet{RequenaTorres2006} that the sputtering of ice mantles of dust grains in shocks may be responsible for the 'rich' chemistry observed in the Galactic centre clouds. 

The interesting question is the degree of processing of gas material and grain mantles in the shock, the survival of molecules in the hot postshock gas \citep{Holdship2017,Palau2017}. The result can be a considerably different chemical composition in the shocked gas than observed in quiescent clouds \citep{Bergin1998}. In order to relate the emission line spectra to the properties of the shocked gas such as gas density and abundances of chemical species, it is necessary to treat comprehensively the dynamic evolution of the gas, the chemistry of the medium, and physical processes of ion and molecule excitation and de-excitation \citep{Flower2015}. 

Different types of shocks can be distinguished depending on the value of the magnetosonic speed in the interstellar gas \citep{Flower2007,Draine2011}. If the shock speed is higher than any signal speed in the shocked medium, the jump (J-type) shock forms. In J-type shocks, physical conditions at shock front change in a discontinuous way, leading to dissipation of the flow kinetic energy in a thin region. As a consequence, high peak temperatures and full dissociation of molecules take place. At low shock speeds, magnetosonic waves precede the shock, and the coupling between ions and neutral gas results in a continuous change in physical parameters of the gas. The continuous (C-type) shocks are formed in this case. Given that H$_2$ molecule is the main coolant of the medium, C-type shocks can exist only up to a limiting value of the shock speed at which the collisional dissociation of H$_2$ (and other molecules) takes place in the hot postshock gas. For typical physical conditions inside dark molecular clouds, the transition from C-type to J-type shock takes place at shock speeds about 40--60~km~s$^{-1}$ \citep{Draine1993,LeBourlot2002}. Models of magnetohydrodynamic non-dissociative shocks in dense molecular clouds have been created by a number of authors, e.g. \citet{Mullan1971,Draine1980,Draine1983,Kaufman1996,Wardle1998,Guillet2007,VanLoo2009,Flower2015}. Numerical models of shock waves usually consider in detail either the gas dynamics, but reduced chemical network is used, or vice versa -- the parametric model of the steady state profile of the shock is used to study in detail chemical evolution of the gas \citep{Holdship2017}. The advantage of a magnetohydrodynamic model over the parametric one is that a large number of physical parameters (e.g. cosmic ray ionization rate, intensity of interstellar background radiation field, dust properties, and etc.) may be varied. Here, we present a magnetohydrodynamic model of C-type shock coupled to a full gas--grain chemical network.

The calculations consist of two steps. In the first step, we compute the chemical composition of a dense cloud with a fixed density. The abundances of chemical species, gas and dust temperatures obtained during this step are used as the initial chemical and physical state of the gas for the shock simulations.

\section{Parameters of the dark cloud}
In this section, physical parameters of the molecular cloud are described. We use the set of 'low-metal' initial abundances of elements except for He, C, and N \citep{Graedel1982}. For He we use a value of 0.09 with respect to hydrogen \citep{Wakelam2008}. For C and N the abundances close to that observed in $\zeta$ Oph diffuse cloud are used \citep{Jenkins2009,Hincelin2011}. The C/O elemental ratio in non-refractory material in dense clouds is not well known, and here we adopt the C/O ratio of 0.5 \citep{Hincelin2011}. The species are assumed to be initially in atomic form except for hydrogen, which is assumed to be molecular. The initial elemental fractional abundances relative to the total H nucleus number density are given in the Table~\ref{table:elabund}, along with the choice of ionization state. All species are assumed to be in the gas phase.

\begin{table}
\caption{Initial elemental abundances with respect to H nucleus number density.}
\begin{tabular}{lc}
\hline
Species & Abundances \\
\hline
H$_2$ & 0.5 \\
He & 0.09 \\
N & $6.2 \times 10^{-5}$ \\
O & $2.8 \times 10^{-4}$ \\
C$^{+}$ & $1.4 \times 10^{-4}$ \\
S$^{+}$ & $8 \times 10^{-8}$ \\
Si$^{+}$ & $8 \times 10^{-9}$ \\
Fe$^{+}$ & $3 \times 10^{-9}$ \\
Na$^{+}$ & $2 \times 10^{-9}$ \\
Mg$^{+}$ & $7 \times 10^{-9}$ \\
Cl$^{+}$ & $10^{-9}$ \\
\hline
\end{tabular}

\small
\medskip
The elemental abundance is the ratio of the number of nuclei both in the gas and in icy mantles of dust grains to the total number of H nuclei, the nuclei locked in the refractory part of the grains are not considered.
\label{table:elabund}
\end{table}

The measured values for the cosmic ray ionization rate in dense interstellar gas lie in the wide range from $10^{-17}$~s$^{-1}$ to values as high as $10^{-15}$~s$^{-1}$ \citep{Dalgarno2006}. A scatter in measured values may be due to details of the measurements, the physical and chemical models used in the analysis of observational data, and may also reflect intrinsic variations of the cosmic ray flux from cloud to cloud. Magnetic field effects can significantly reduce the cosmic ray ionisation in dense cloud cores \citep{Padovani2011}.

The dispersion of turbulent velocities (micro-turbulence speed) and gas velocity gradient are necessary parameters for radiative transfer calculations. The micro-turbulence speed determines the excess in line width over the thermal value while the gas velocity gradient determines the length of the region where molecular radiation is coupled to the ambient gas. Starless cores in molecular clouds -- sites of low-mass star formation -- present spectra of core-tracing species that have close-to-thermal line widths \citep{Tafalla2004,Andre2014}. Here, the dispersion of turbulent velocities is taken equal to the sound speed in the gas at 10~K -- 0.2~km~s$^{-1}$. Molecular clouds are not rigorously characterized by large-scale systematic motion, as required for the large velocity gradient approximation to be valid. However, the characteristic value of the velocity gradient can be estimated as the ratio of cloud line width (in velocity units) and cloud radius. From this, one can estimate the velocity gradient of the order of $1$~km~s$^{-1}$~pc$^{-1}$ \citep{Goldsmith2001}.

The nature and structure of shock waves travelling through molecular clouds are strongly dependent upon the strength of magnetic field $B_0$ \citep{Draine1980}. According to the analysis of the data on magnetic field strength in molecular clouds by \citet{Crutcher1999}, an approximate empirical relation holds between line-of-sight magnetic field strength and gas density $B_{\rm{los}} = \beta n_{\rm{H,tot}}^{1/2}$, where $n_{\rm{H, tot}}$ is the total hydrogen nuclei number density, $\beta \approx 0.7$~$\mu$G~cm$^{3/2}$. We assume that $B_0/B_{\rm{los}} \simeq$ 1.5--2.

A 'classical' single-size grain model is considered. The grains, made of silicate material, are assumed to be spherical particles with a radius of 0.1~$\mu$m and internal density of 3.5~g~cm$^{-3}$. The dust--gas mass ratio is taken equal to 0.01. The grains are initially bare. We neglect the change in the grain radius by freeze out of gas-phase molecules onto dust grains. Optionally, our computer code is able to treat polycyclic aromatic hydrocarbon (PAH) molecules, but no PAH molecules are considered in our 'standard' model. Optical properties for silicate and carbonaceous spheres are used from \citep{Draine1984,Laor1993,Li2001,Weingartner2001} and are available at website of Prof. B.T.~Draine\footnote{\url{http://www.astro.princeton.edu/~draine/index.html}}.

The initial ortho-/para-H$_2$ ratio is taken to be 1 as a representative value of dark molecular clouds at young ages, $t < 1$~Myr \citep{Pagani2013}. The chemistries of ortho- and para-H$_2$ are not distinguished in our model. 

The physical parameters of our 'standard' model are given in the Table~\ref{table:modelparam}. The description of the simulations of gas-phase and grain surface chemistries, calculations of level populations of ions and molecules, heating and cooling processes, and shock structure are given in the appendices \ref{app_chemistry}, \ref{app_levpop}, \ref{app_heatcool}, and \ref{app_shock}, respectively. The list of chemical reactions of collisional dissociation of species is given in the appendix \ref{app:colldissreactions}.

\begin{table}
\caption{Parameters of the dark cloud.}
\begin{tabular}{ll}
\hline
Parameter & Value \\
\hline
Gas density, $n_{\rm{H, tot}}$ & $2 \times 10^4$~cm$^{-3}$\\
Magnetic field strength, $B_0$ & 0.15~mG \\
Visual extinction, $A_{\rm{V}}$ & 10 \\
Micro-turbulence speed, $v_{\rm{turb}}$ & 0.2~km~s$^{-1}$ \\
Velocity gradient & 1~km~s$^{-1}$~pc$^{-1}$ \\
Cosmic ray ionization rate, $\zeta$ & $3 \times 10^{-17}$~s$^{-1}$ \\
Scaling factor of local interstellar radiation, $G_0$ & 1 \\
Dust--gas mass ratio & 0.01 \\
Grain radius, $a$ & 0.1~$\mu$m \\
Grain material density & 3.5~g~cm$^{-3}$ \\
Grain surface area density & 4.8$\times 10^{-22}$~cm$^{2}$~per~H \\
Ortho-/para-H$_2$ ratio & 1 \\ 
\hline
\end{tabular}
\label{table:modelparam}
\end{table}

\section{Results}
\subsection{Chemical evolution of the dark cloud}
The objectives of the modelling of chemical evolution of a static dark cloud are: (i) verification of our chemical model; (ii) evaluation of the chemical composition of the gas before the shock wave propagates through it. In simulations of chemical evolution of a static dark cloud, a simple zero-dimensional model is considered in which density remains fixed as the chemistry progresses from initial specimen abundances. In this section the main results of our calculations are presented, and the comparison is made with the results of other workers.
 
\subsubsection{General results}

\begin{figure}
\includegraphics[width=180mm]{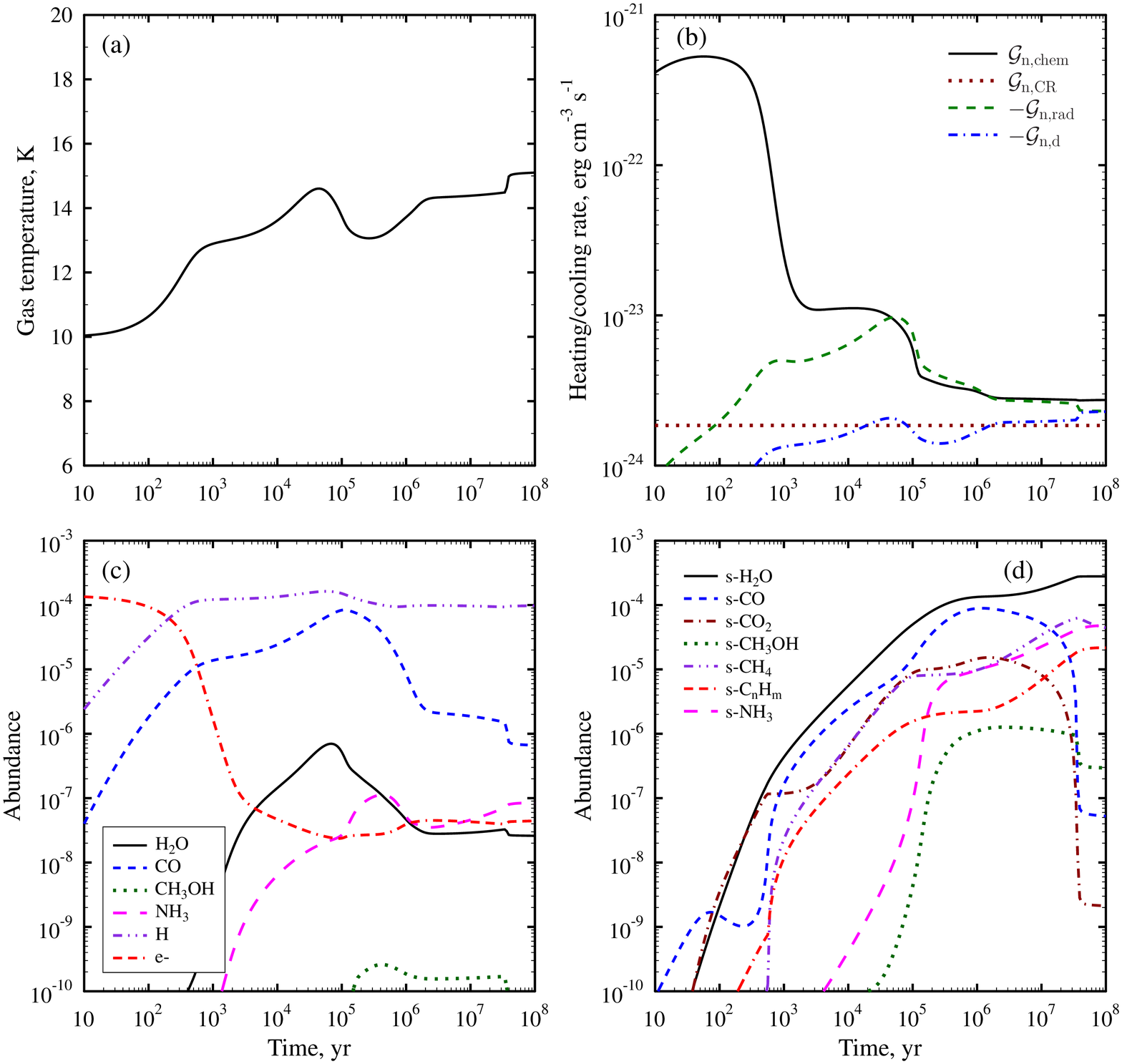}
\caption{Chemical and thermal evolution of the static cloud: (a) gas temperature; (b) rates of main heating and cooling mechanisms of the gas, the term $\mathcal{G}_{\rm{n, rad}}$ includes cooling by ions CI, CII, OI and molecules H$_2$, CO and H$_2$O; (c) abundances of gas species relative to hydrogen nuclei; (d) abundances of grain mantle species, s-X denotes species X in icy mantles, C$_n$H$_m$ denotes all hydrocarbon molecules with $n \geq 2$.}
\label{fig_chem_therm_ev_2e4}
\end{figure}

The temperature of the neutral gas component is shown on the Fig. \ref{fig_chem_therm_ev_2e4}a. During the evolution of the dark cloud all gas components have close temperatures. The gas reaches a thermal equilibrium after about $10^3$~yr and gas temperature remains at relatively constant level of about 13--15~K. The main heating mechanism of the gas is the cosmic ray driven chemistry, see Fig. \ref{fig_chem_therm_ev_2e4}b. The main cooling mechanisms are cooling via molecular and atomic line emission, and by gas--dust collisions. At evolutionary times $t > 10^5$~yr, heavy species become adsorbed on dust grains (see Fig. \ref{fig_chem_therm_ev_2e4}c). But total gas-phase depletion of heavy species does not take place due to reactive desorption mechanism. At gas densities $n_{\rm{H,tot}} \gtrsim 10^4$~cm$^{−3}$, the gas and dust start to couple thermally via collisions and molecular depletion effect on the gas temperature diminishes, see also \citet{Goldsmith2001}. 

The main ice mantle constituent of interstellar grains is H$_2$O ice (Fig. \ref{fig_chem_therm_ev_2e4}d). According to our simulations, water on the grain surface is mainly formed in the reaction between H atom and hydroxyl OH, with some contribution of other channels. In our model, the contribution of CO$_2$ to icy mantles is approximately 10--20 per cent that of the H$_2$O ice at 10$^5$--10$^6$~yr. Carbon dioxide forms on the grain surface via reactions CO+OH and H+HOCO (HOCO in turn is also produced in reaction CO+OH). The dominance of one or the other reaction channel in CO$_2$ formation strongly depends on the adopted parameters such as thickness and height of activation barriers of chemical reactions. The abundance of methanol is relatively low, 0.1--1 per cent that of water abundance at 10$^5$--10$^6$~yr. Hydrogen atoms adsorbed on dust grains participate in numerous surface reactions. As a result, hydrogen molecule formation via direct association of hydrogen atoms is negligible except at very late evolutionary times $t > 3 \times 10^7$~yr. According to our simulations, hydrogen molecule is produced through hydrogen abstraction reactions on grain surface from molecules HCO, H$_2$CO, HNO, and other species, see also \citet{Tielens1982,Hasegawa1992}. The abundances of species in ice mantles of grains found in our simulations are in reasonable agreement with astronomical ice observations \citep{Boogert2015}.

The temperature of dust grains is equal to 9.3~K according to our calculations. At $A_{\rm{V}} \gtrsim 3$, dust grains are mainly heated by interstellar radiation field at infrared wavelengths. Most of dust grains are either neutral or have the charge $-1$. According to our simulations, the photoelectric emission by cosmic ray induced UV radiation and collisional attachment of ions and electrons are all important in grain charging as ionization fraction has reached values of 10$^{-7}$--10$^{-8}$. Analogous results were found by \citet{Ivlev2015}. The effect of cosmic ray induced UV radiation field on grain charge decreases with increasing gas density \citep{Guillet2007}.  

\subsubsection{Comparison with NAUTILUS model}

\begin{figure}
\includegraphics[width=110mm]{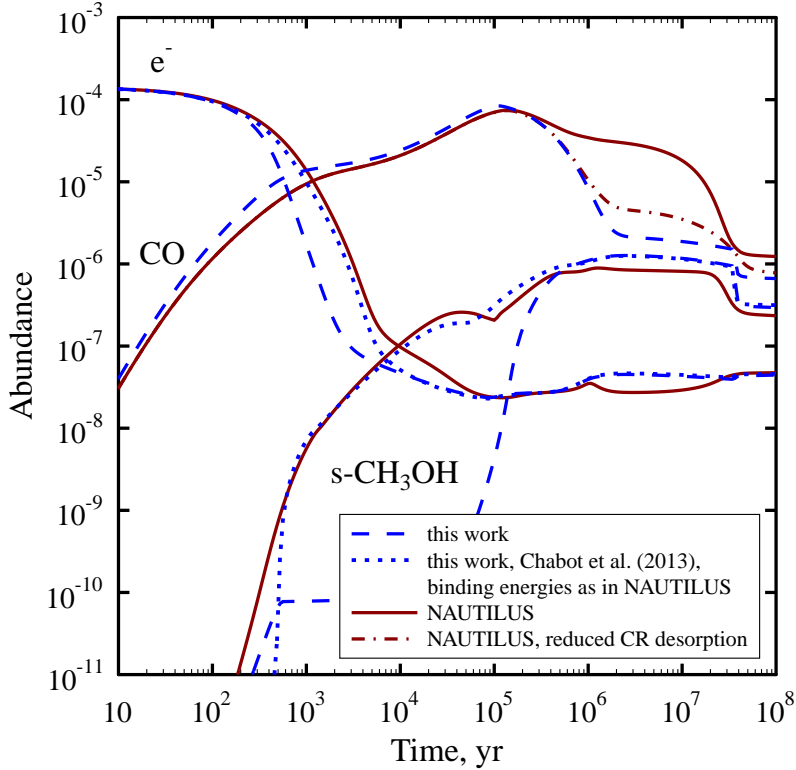}
\caption{Comparison of our simulation results with NAUTILUS results. The results of simulations are also shown with updated chemical network according to \citet{Chabot2013} and using the data on specimen binding energies as in NAUTILUS model. The results of NAUTILUS model with direct cosmic ray desorption reduced by an order of magnitude are shown.} 
\label{fig_comp_nautilus}
\end{figure}

The comparison is made between the results of our calculations and the NAUTILUS code \citep{Ruaud2016}, see Fig. \ref{fig_comp_nautilus}. In NAUTILUS simulations, we set identical to our model the initial elemental abundances (Table \ref{table:elabund}), parameters of grain surface chemistry (Table \ref{table:chemistryparam}), gas temperature evolution with time (Fig. \ref{fig_chem_therm_ev_2e4}a). The difference between the results of our simulations and that of NAUTILUS code is no higher than an order of magnitude for most simple species. Note that different data on binding energies and different databases for gas-phase chemical reactions are used in our calculations and in NAUTILUS code.

The method of direct cosmic ray desorption of adsorbed species by \citet{Hasegawa1993a} is incorporated in NAUTILUS code, whilst the method by \citet{Roberts2007} is used in our model, see appendix \ref{app_chemistry}. The cosmic ray desorption of volatile molecules such as CO is very efficient according to NAUTILUS simulations, and the abundance of CO in the gas phase stays at high level. If the cosmic ray desorption is reduced by an order of magnitude, the results become much closer, see Fig. \ref{fig_comp_nautilus}. According to our results, the decrease of the ionization fraction of the gas starts at early evolutionary times compared with NAUTILUS results. At about 10$^2$--10$^3$~yr, large unsaturated carbon-chain molecules C$_n$ and C$_n$H and their anions are produced in significant amounts in our model, and the gas is effectively neutralized through reactions:

\begin{equation}
\begin{array}{l}
\displaystyle
\rm{C_n}^{-} + \rm{C}^{+} \to \rm{C_n} + \rm{C}, \quad \rm{C_nH}^{-} + \rm{C}^{+} \to \rm{C_nH} + \rm{C}.
\end{array}
\label{eq_gas_neutr_hyca}
\end{equation}

\noindent
\citet{Chabot2013} provided a set of branching ratios for the reactions involving carbon-chain species. For the reaction $\rm{C_n}^{-}+\rm{C}^{+}$, internal energy of intermediate complexes is high, and three fragment channels are dominating. These branching ratios are so far not taken into account in the UDfA chemical network but are included in the latest version of the KIDA network \citep{Wakelam2015} that is used in NAUTILUS. New branching ratios prevent formation of large carbon-chain molecules at early times \citep{Chabot2013}. Indeed, our results on electron abundance become very close to NAUTILUS results when we update our chemical network according to data by \citet{Chabot2013}, see Fig. \ref{fig_comp_nautilus}. 

\citet{Vasyunin2004} analysed the influence of errors in the rate constants of gas-phase chemical reactions on the calculated specimen abundances. They found that errors in the abundances of simple species lie within 0.5--1 order of magnitude. \citet{Wakelam2006} found similar uncertainties and discussed the differences between two wide used gas-phase chemical networks. Recently, \citet{Penteado2017} presented a systematic study of the effect of uncertainties in the binding energies on abundances of chemical species. They found that there is a large variation in the abundances of ice species when binding energies are varied within their errors. Our results on CH$_3$OH abundance in icy mantles become close to those by NAUTILUS model, if the same data on binding energies are used in both models, see Fig. \ref{fig_comp_nautilus}.

\subsection{Shock model results}
According to the {\it Herschel} Gould Belt survey studies of nearby star-forming clouds, the typical lifetime of starless cloud cores with density $\sim 10^4-10^5$~cm$^{-3}$ is about 10$^6$~yr on average \citep{Andre2014}. Moreover, best agreement between observed and modelled specimen abundances in cloud cores is most often achieved at 'early' times of 10$^4$--10$^6$~yr \citep{Wakelam2006,Penteado2017}. For the shock wave modelling, the gas chemical composition and gas temperature at 0.5~Myr are chosen.
 
\subsubsection{Comparison with the shock model by \citet{Flower2015}}

\begin{figure}
\includegraphics[width=180mm]{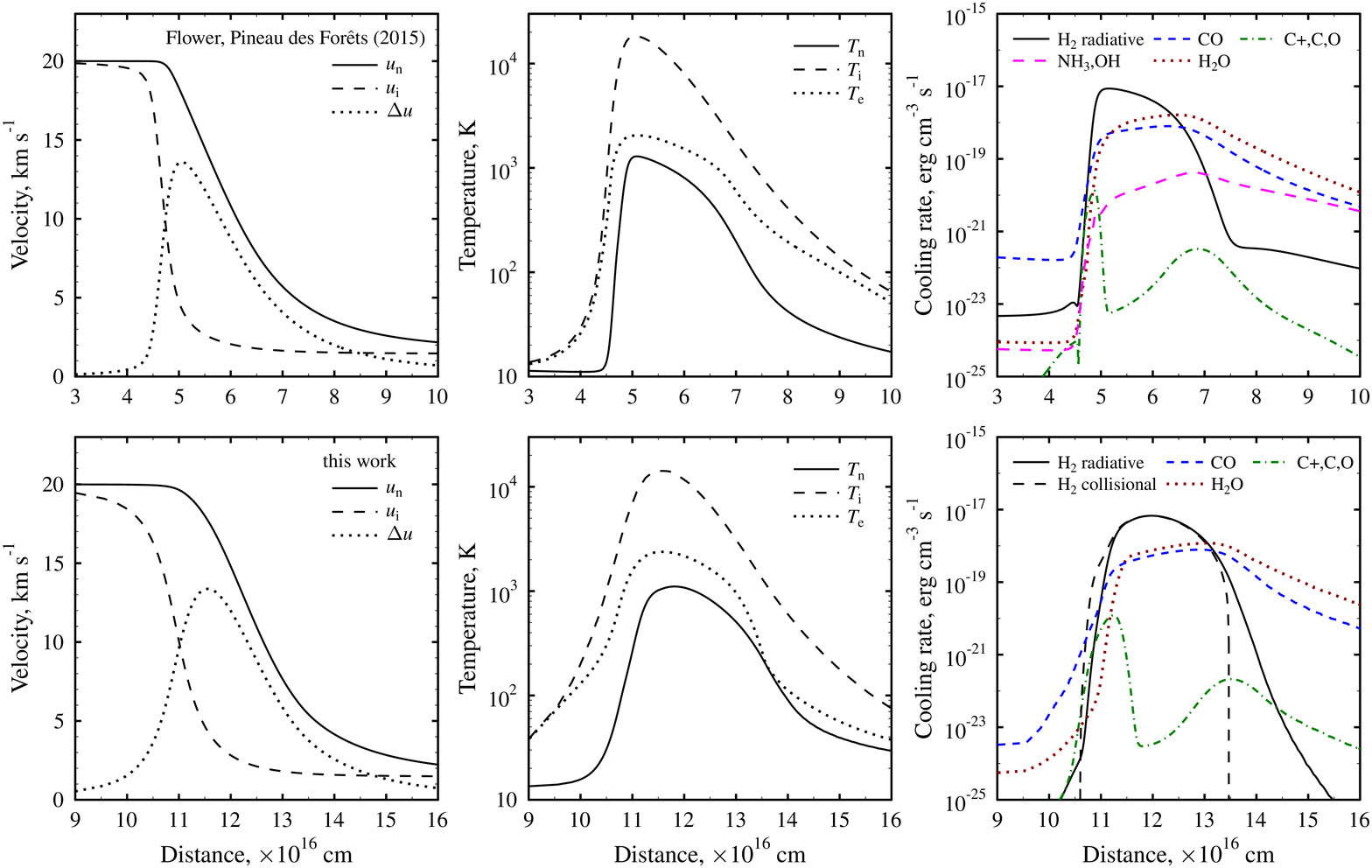}
\caption{Comparison between results of our calculations, lower panels, and the C-type shock structure calculated by the model by \citet{Flower2015}, upper panels, for a shock velocity $u_{\rm{s}}$ = 20~km~s$^{-1}$. The integration of differential equations starts at the zero point of the $z$-axis. The rates of thermal energy transfer in collisions from the gas to molecules and ions are shown ('collisional'), the net rate of radiative energy loss is also given for H$_2$ molecule ('radiative').}
\label{fig_flower_comp}
\end{figure}

A comparison is made between the results of our simulations and the results obtained using the code \verb"mhd_vode" by \citet{Flower2015}. The similar specimen abundances are used at the start of both simulations. Fig. \ref{fig_flower_comp} shows velocities and temperatures of gas components, and gas cooling rates due to emission in molecular and atomic lines. The general behaviour of physical parameters is similar in our model and that by \citet{Flower2015} -- both models provide approximately the same shock width, maximal temperatures of gas components, and the same maximal velocity difference between ion and neutral fluids. The shock model by \citet{Flower2015} does not consider grain surface chemistry and includes simpler gas-phase chemical network than our model -- the influence of expanded chemistry on shock structure is minimal.

The rates of collisional energy transfer from the gas to the molecules and ions are shown in the Fig. \ref{fig_flower_comp}, the net rate of radiative energy loss is also given for H$_2$ molecule (assuming that radiative transitions are optically thin). The main gas coolant in hot shocked gas is H$_2$ molecule. As the shocked gas cools, the contribution of other molecules to the cooling process becomes significant. The gas cooling by molecules OH, NH$_3$ and CH$_3$OH is not yet taken into account in our model. Cooling by these species may be significant in the case of elevated abundances \citep{Flower2010b}. In the postshock region, rotational levels of ground vibrational state of H$_2$ are overpopulated due to slow de-excitation rates of these levels. In this case, the collisional energy transfer from the gas to the H$_2$ molecule changes the sign, and collisions of gas species with H$_2$ heat the gas, see Fig. \ref{fig_flower_comp}. Our model produces low rate of radiative energy loss by H$_2$ in the cold gas where H$_2$ molecules can be vibrationally excited only by formation processes -- this effect is not taken into account in our model.

\subsubsection{Chemical evolution of the shocked gas}

\begin{figure}
\includegraphics[width=170mm]{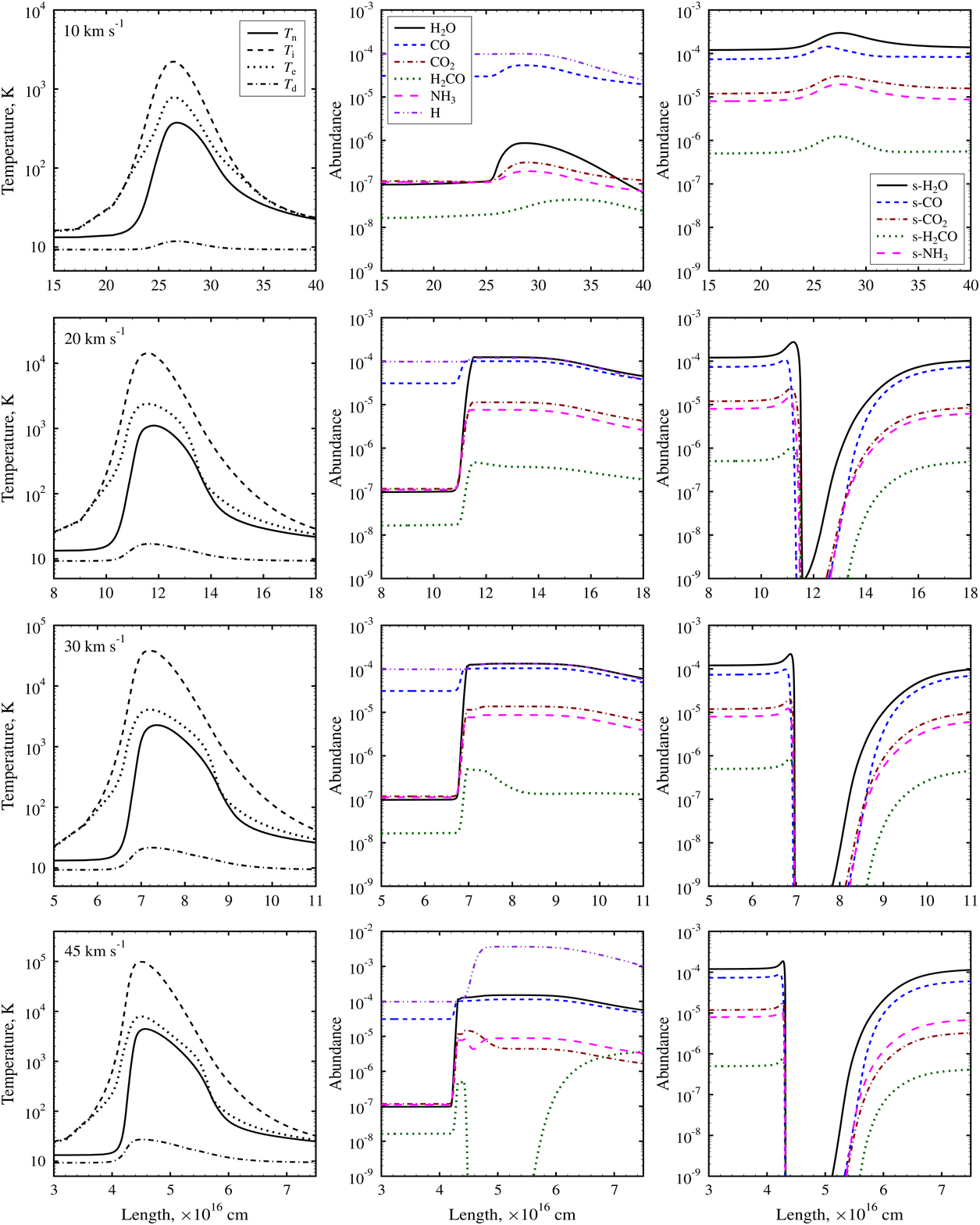}
\caption{Shock structure and evolution of abundances of simple species. Results are shown for four shock velocities: 10, 20, 30, 45~km~s$^{-1}$. Each row of graphs corresponds to the shock velocity shown on the corner of the first graph in the row.}
\label{fig_shock_chem_2e4}
\end{figure}

Fig. \ref{fig_shock_chem_2e4} shows shock structure and evolution of abundances of simple molecules that are often considered as shock tracers. There is an increase of abundances of icy mantle species at the beginning of the shock before sputtering starts -- the grain velocity decreases and their number density increases in the magnetic precursor as the grains are (partially) coupled to the ion fluid. The sputtering of grain mantles starts when the gas--grain relative speed reaches about 5~km~s$^{-1}$. Thus, the threshold shock speed for grain mantle sputtering is about 10~km~s$^{-1}$. The main sputtering projectiles are heavy species He and CO at low shock speeds. These results are similar to those by \citet{JimenezSerra2008,VanLoo2013}. Molecules released from grain mantles are chemically processed in the hot postshock gas. As the gas cools, high abundances of species produced in the shock persist in the postshock gas until the time-scale for the individual molecule to deplete onto dust grains. The simulation results on abundances of simple species are in agreement with the findings by other workers, see e.g. \citet{Bergin1998,Charnley2000,Viti2011,Flower2012}. 

\begin{figure}
\includegraphics[width=125mm]{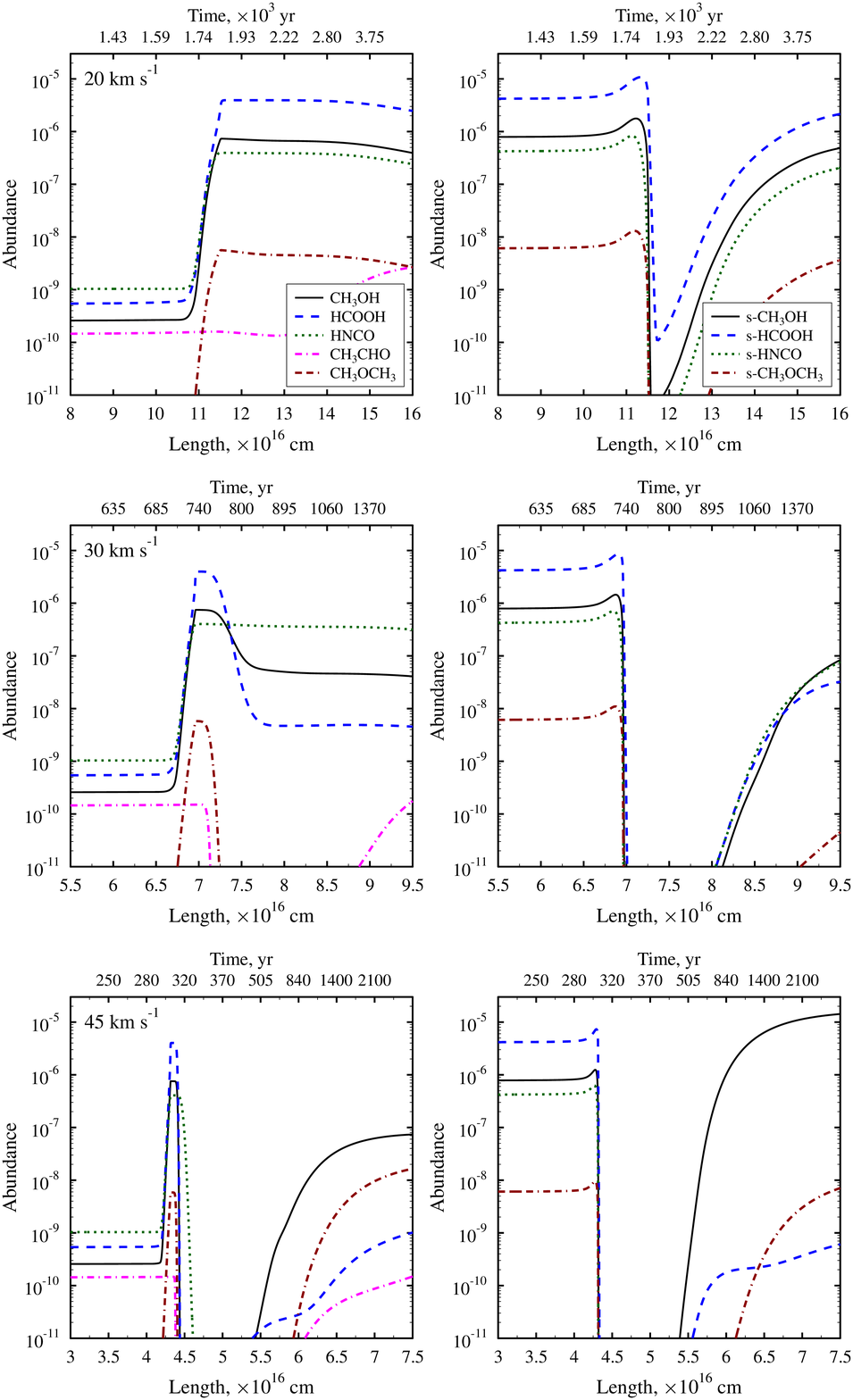}
\caption{Evolution of abundances of some COMs in the shock: methanol CH$_3$OH, formic acid HCOOH, isocyanic acid HNCO, acetaldehyde CH$_3$CHO, dimethyl ether CH$_3$OCH$_3$. Results are shown for shock velocities 20, 30 and 45~km~s$^{-1}$. }
\label{fig_shock_COMs}
\end{figure}

Fig. \ref{fig_shock_COMs} shows the evolution of abundances of some COMs in the shock wave. Species that are produced in 'nonenergetic' atom addition reactions (e.g. CH$_3$OH, CH$_3$OCH$_3$) are abundant in icy mantles in the preshock gas. Abundances of many other complex species (e.g. HCOOCH$_3$, C$_2$H$_5$OH) are low, as radical--radical association reactions that produce such species are inefficient at low dust temperatures. In shock, the increase of ion--neutral drift velocity is rapid, and gas reaches the maximum temperature soon after the sputtering. Hence, the sputtering of grain mantles takes place in the region close to the temperature peaks of neutral gas and ions. At high shock speeds, molecules are destroyed in the hot shocked gas via reactions with H atoms and collisional dissociation reactions. The survival time of complex molecules in the hot shocked gas is low -- of the order of dozens of years, see Fig. \ref{fig_shock_COMs}.

At high shock speeds, there is high abundance of atomic hydrogen in the gas produced in collisional dissociation reactions. It triggers formation of hydrocarbon molecules on the grain surface. At shock speed 45 km s$^{-1}$, the abundance of methanol in icy mantles of dust grains in the cool postshock gas equals to 10--15 per cent relative to water ice, while in the preshock gas it constitutes only about 0.5 per cent, see Fig. \ref{fig_shock_COMs}. According to observational data, the abundance of methanol ice (relative to H$_2$O ice) ranges from upper limits of no more than a few percent toward dense molecular clouds to substantial levels of up to 30 per cent toward a few young stellar objects, e.g. RAFGL7009S and W~33A \citep{Dartois1999,Pontoppidan2003}. The methanol production in the cooling postshock gas is one of the possible explanations of high abundance variations of methanol ice observed in astronomical sources. Methanol is reformed in the gas phase via production on the grain surface followed by the desorption into the gas and, with a minor contribution, via gas-phase reactions. The efficiency of reactive desorption mechanism is a key parameter that controls methanol re-formation in the gas phase in the cooling postshock region. At shock speed 45 km s$^{-1}$, the peak abundance of gas-phase methanol in the postshock gas is $8 \times 10^{-8}$ at standard model parameters, and is about $7 \times 10^{-9}$ at efficiency of reactive desorption of $f$ = 0.001 -- an order of magnitude lower than our standard value.

COMs are effectively produced in the gas phase in the postshock region. At about 10$^4$ years after the passage of high speed shock (45 km s$^{-1}$), the abundance relative to H nuclei of methyl formate HCOOCH$_3$ (both in the gas and in icy mantles) is about $4 \times 10^{-10}$ that is almost three orders of magnitude higher than in the preshock gas. The analogous effect is seen for ethanol C$_2$H$_5$OH, but the abundance of ethanol reaches low values of about 10$^{-11}$ in our model. The abundance of acetaldehyde CH$_3$CHO in icy grain mantles is low as it reacts rapidly with hydrogen atoms. In postshock region, acetaldehyde is effectively produced in the gas phase, it abundance reaches values of about 10$^{-9}$ -- an order of magnitude higher than in the preshock gas. The main parent species in the gas-phase synthesis of COMs are H$_2$CO, CH$_3$OH, C$_2$H$_4$ and radicals CH$_3$, CH$_3$O, C$_2$H$_5$.

\section{Discussion}
The reactions of 'nonenergetic' atom addition on the grain surface play a main role in dark cloud chemistry as adsorbed species are not able to cross large reaction barriers \citep{Charnley2008,Fedoseev2017}. Parameters that have a strong influence on the simulated grain mantle composition are the adopted binding and diffusion energies of species and activation energies of grain surface reactions \citep{Taquet2012,Penteado2017}. The chemical composition of icy mantles in dark clouds strongly depends on the grain temperature, cosmic ray ionization rate, gas density and evolution stages experienced by the gas prior to significant grain mantle formation. The problem of chemical evolution of a particular dark cloud and shock propagation through it must be considered jointly in order to make reasonable predictions on specimen abundances and molecular line intensities.

Most grain surface reactions are exoergic. Part of the energy released in a reaction will be immediately transferred to the ice mantle of dust grain, other part is transformed to the kinetic energy and internal (electronic and ro-vibrational) excitation of reaction products \citep{Lamberts2014,Fredon2017}. We have not included dust heating by chemical reactions. Dust heating by this mechanism is negligibly small at dark cloud conditions. However, an explosive release of the chemical energy stored in icy mantles as free radicals may take place as the dust temperature rises in the shock -- this mechanism was considered by \citet{Shen2004} in the study of cosmic ray induced explosive chemical desorption in dense clouds. Quite possible that this mechanism can lead to the liberation of volatile species from icy mantles to the gas phase before sputtering in the shock. At the model parameters in question, the effect of dust heating on the shock chemistry is small -- the dust temperature is increased up to 15 K in the shocked gas before grain mantle sputtering, but the gas passes quickly through this region. The possible influence of dust heating on the grain surface chemistry in the shock is beyond the scope of the present paper.

Observations of molecular outflows of protostars indicate that molecules show two different kinds of profiles, with CH$_3$OH, NH$_3$ and other species emitting only at relatively low outflow velocities, whereas H$_2$O shows bright emission even at the highest velocities \citep{Codella2010,Gomez-Ruiz2016,Holdship2016}. It is explained by chemical modelling -- molecules that have destruction reactions with low activation energy are destroyed in hot gas \citep{Viti2011}. The discussion on the behaviour of simple molecules in C-type shocks and their potential to be a shock tracer was given by \citet{Holdship2017}. According to our simulations, the time-scale for molecule survival in the high-speed shock is low -- of the order of dozens of years. Recently, \citet{Palau2017} studied the evolution of COMs using the parametric shock model by \citet{JimenezSerra2008}. They found that COMs can survive long enough after the passage of a shock. Our results imply that more strict constraints must be put on the physical parameters for the shock regions where COMs are observed -- shock velocity and gas density must be low enough to allow COMs survive in the hot shocked gas. Other possibility is that observed COM's emission comes from a postshock region where these molecules have been reformed in the gas \citep{Codella2015,Palau2017}. It is likely that a fraction of molecules could be destroyed in the sputtering process, but we do not consider this effect in our model, see discussion by \citet{Suutarinen2014}. 

The COMs destroyed in the hot shocked gas could be reformed in the postshock region. The high abundance of simple species and radicals that are produced in the sputtering of icy mantles and collisional dissociation reactions is one of the factors that promote COM production in the postshock gas. The abundance of H atoms is other important factor controlling the chemistry in the interstellar gas \citep{Cuppen2009,Wakelam2010}. As the gas cools, H atoms drop out the gas through the adsorption on dust grains. It is shown that the high abundance of H atoms in the cooling postshock gas may trigger formation of methanol on the grain surface. Indeed, as was shown by \citet{Cuppen2009}, the increase of H/CO ratio in the gas in dark molecular clouds shifts the grain surface chemistry to the formation of complex hydrocarbon molecules. In the postshock region, gas-phase methanol is mainly produced via reactive desorption mechanism. The observations of methanol in the cold gas that has experienced the shock passage in the recent history can be used to quantify the magnitude and importance of reactive desorption mechanism. The methanol is the parent specimen in the production of other COMs and the gas-phase abundance of such species strongly depends on methanol abundance. 

One of the shortcomings of our model is the simple dust model. The size distribution of grains has significant effect on the chemical evolution of the dark cloud \citep{Iqbal2018arXiv} and on the shock structure and the gas-phase chemistry in the shock \citep{VanLoo2009,Flower2012}. Grain shattering in shocked gas may produce numerous small grain fragments, which increase the total dust grain surface area. The effect is that C-type shocks become shorter and warmer, which in turn affects the chemistry and molecular emission \citep{Anderl2013}. Grain shattering and its feedback onto the dynamics of C-type shocks are found to be significant at densities higher than about $10^5$~cm$^{-3}$ \citep{Guillet2011}. 

The chemical network used in our model is incomplete for proper modelling of chemistry of COMs and extensions of gas-phase and grain surface chemistries must be used in future work \citep{Garrod2013b,Taquet2016,Fedoseev2017}.

\section{Conclusions}
The evolution of abundances of COMs in C-type shock wave is considered. The main results of the paper are summarized below:

(i) The sputtering of grain mantles takes place in the shock region close to the peak of neutral gas temperature. As a result, time-scale for molecule survival in the high-velocity shock is low  -- of the order of dozens of years.

(ii) For high-speed shocks, $u_{\rm{s}} \gtrsim 40-45$~km~s$^{-1}$ at preshock gas density $n_{\rm{H,tot}} = 2 \times 10^4$ cm$^{-3}$, the abundance of H atoms and radicals in the postshock gas is relatively high, that affects the gas-phase and grain surface chemistries. The efficient methanol production on the surface of dust grains in the cool postshock gas may be one of the reasons of high abundance of methanol ice observed toward some young stellar objects. 

(iii) Gas-phase methanol is re-formed in the postshock gas via reactive desorption mechanism. The efficiency of reactive desorption is a key parameter that determines the gas-phase abundance of methanol and other complex species that are produced via it. 

(iv) At high shock speeds, the postshock abundances of some COMs such as methyl formate may be much higher than in the preshock gas due to efficient gas-phase production.

A comparison has been made between our simulations and results obtained with other publicly available codes: NAUTILUS for modelling chemistry of dark clouds \citep{Ruaud2016} and \verb"mhd_vode" for modelling shocks \citep{Flower2015}. The difference between the results of our simulations and those of NAUTILUS code is not larger than an order of magnitude for most simple species. There is a good agreement between our results and that of the \verb"mhd_vode" code -- both codes predict similar profiles of gas temperature and velocity, and similar shock widths. 

The shock model presented can be employed to interpret observational data on molecular emission from outflows powered by protostars, to study formation of biologically relevant molecules in shock regions, and can be also used for modelling of cosmic masers. The influence of various physical parameters on the chemistry and molecular emission of the shocked gas can be studied using this model. 

\section*{Acknowledgements} 
This work was supported by the RSF grant 16-12-10225.

\bibliography{../../interstellar_medium_references,../../chemistry_references}
\clearpage

\appendix
\section{Chemistry}
\label{app_chemistry}
\subsection{Gas phase chemistry}
The gas-phase chemical network used here is based on the UMIST Database for Astrochemistry (UDfA), 2012 edition \citep{McElroy2013}\footnote{\url{http://udfa.ajmarkwick.net/}}. The species which contain fluorine, F, and phosphorus, P, are removed in order to reduce computation time. The species, that are not included in the current edition of the UDfA network -- CH$_3$O, CH$_2$OH, HC$_2$O, CH$_3$CO, HOCO, CH$_3$OCH$_2$ -- are added to the network, as we are interested in the production of methanol and some other organic species on the surface of dust grains. Reactions of new species with abundant ions are added to our chemical network \citep{Bacmann2016}. The gas-phase chemistry is supplemented with the list of neutral--neutral reactions published by \citet{Palau2017} and collisional dissociation reactions (see appendix \ref{app:colldissreactions}). Thus, our gas-phase network consists of reactions involving 430 species composed of the elements H, He, C, N, O, Na, Mg, Si, S, Cl, Fe. Unimolecular and bimolecular reactions are considered.

\subsubsection{Main equations} \label{species_data} 
For two-body reactions involving species $j$ and $l$ of the same gas component (neutrals or ions), the rate coefficient is calculated by the modified Arrhenius formula \citep{McElroy2013}:

\begin{equation}
k_{\rm{jl}} = \alpha \left(\frac{T}{300}\right)^{\beta} {\rm{exp}} \left(-\frac{\gamma}{T} \right) \ \left[{\rm{cm^3 s^{-1}}}\right],
\label{arrhenius_formula}
\end{equation}

\noindent
where $\alpha$, $\beta$ and $\gamma$ are reaction parameters, $T$ stands for neutral gas temperature when both reactants are neutrals and for ion temperature when both reactants are ions, $T$ is measured in K. For reactions involving electrons and regardless of the type of a second reactant, the equation (\ref{arrhenius_formula}) is used with the electron temperature.

For direct cosmic ray ionisation reactions, the rate coefficient is given by

\begin{equation}
k_{\rm{j}} = \alpha \frac{\zeta}{\zeta_0} \ \left[{\rm{s^{-1}}}\right],
\end{equation}

\noindent
where $\zeta$ is the cosmic ray ionization rate, $\zeta_0$ is the standard value of the parameter, here $\zeta_0 = 1.36 \times 10^{-17}$~s$^{-1}$ \citep{McElroy2013}. For cosmic ray induced photo-reactions the rate coefficient is 

\begin{equation}
k_{\rm{j}} = \alpha \left(\frac{T}{300}\right)^{\beta} \frac{\gamma}{1 - \omega} \frac{\zeta}{\zeta_0} \frac{2n_{\rm{H_2}}}{n_{\rm{H, tot}}} \ \left[{\rm{s^{-1}}}\right],
\label{CRphoto_reaction_rate}
\end{equation}

\noindent
where $n_{\rm{H_2}}$ is the number density of hydrogen molecules, the dust grain albedo in the ultraviolet $\omega$ is taken equal to 0.5.

For photo-reactions induced by interstellar ultraviolet (UV) radiation field, the rate coefficient is parametrized as

\begin{equation}
k_{\rm{j}} = G_0 \alpha \, {\rm{exp}} \left( -\gamma A_{\rm{V}} \right) \ \left[{\rm{s^{-1}}}\right],
\end{equation}

\noindent
where $A_{\rm{V}}$ is the visual extinction, $G_0$ is the scaling factor. The unshielded photo-reaction rates $\alpha$ given in the UDfA chemical network are calculated for the 'standard' interstellar UV radiation field given by \citet{Draine1978}, for this field and below $G_0 = 1$. Here, we don't take into account the decrease of photo-dissociation rates of H$_2$, CO, N$_2$ molecules due to self-shielding. 

At temperatures higher than the maximal temperature given in the database for a specific reaction, the reaction rate is taken equal to the rate at maximal temperature. The exception is the reactions of collisional dissociation of chemical species. The reaction rates are extrapolated at temperatures lower than minimal temperature if the reaction rate decreases with decreasing temperature.

\subsubsection{Ion--neutral reactions}
Substantial difference in the flow velocities of charged and neutral fluids may help to overcome the energy barrier in endothermic ion--neutral reactions \citep{Draine1986a}. The ions and neutrals are assumed to have temperatures $T_{\rm{i}}$ and $T_{\rm{n}}$ and mean velocities $u_{\rm{i}}$ and $u_{\rm{n}}$, respectively. Let $u_{\rm{in}}$ to be the ion--neutral drift velocity, $u_{\rm{in}} = u_{\rm{i}} - u_{\rm{n}}$. In calculations of reaction rates, the equation (\ref{arrhenius_formula}) is used with an 'effective' temperature $T_{\rm{eff}}$ \citep{Flower1985}: 

\begin{equation}
\displaystyle
T_{\rm{eff}} = T_{\rm{r}} + \frac{m_{\rm{in}} u_{\rm{in}}^2}{3 k_{\rm{B}}}, \\ [15pt]
\label{efftemp}
\end{equation} 

\noindent
where $m_{\rm{in}}$ is the reduced mass of reactants and $T_{\rm{r}}$ is their weighted kinetic temperature. The discussion on the accuracy of such approximation was given by \citet{Draine1986a}.

\subsection{Gas--grain interactions}
\subsubsection{Adsorption}
The accretion rate $k_{\rm{a, j}}$ of species $j$ onto the surface of dust grains is:

\begin{equation}
\displaystyle
k_{\rm{a, j}} = s_{\rm{j}} \sigma_{\rm{g}} \overline{v}_{\rm{th, j}} n_{\rm{g}} \ \left[{\rm{s^{-1}}}\right], \ \overline{v}_{\rm{th, j}} = \left( \frac{8 k_{\rm{B}} T}{\pi m_{\rm{j}}} \right)^{1/2},
\end{equation}

\noindent
where $s_{\rm{j}}$ is the sticking coefficient, $\sigma_{\rm{g}} = \pi a^2$ is the geometrical cross section of a dust grain with radius $a$, $\overline{v}_{\rm{th, j}}$ is the average thermal velocity of species $j$ with mass $m_{\rm{j}}$, $T$ is the kinetic temperature of incident particles, and $n_{\rm{g}}$ is the number density of dust grains. For H and H$_2$, we use the expressions for sticking probability dependence on gas temperature given by \citet{Matar2010}. Sticking coefficients of other species are assumed equal to 1 at low gas and dust temperatures. We assume that the dependence on gas temperature of sticking probability of heavy species has the same form as for H$_2$ molecule, but the correction for particle mass is made. There is a critical value of gas temperature $T_{\rm{0,j}}$ above which the sticking probability of specimen $j$ starts to decline, for H$_2$ molecule $T_{\rm{0,H_2}}=87$~K \citep{Matar2010}. Here we crudely assume $T_{\rm{0,j}} = T_{\rm{0,H_2}} m_{\rm{j}}/m_{\rm{H_2}}$. The sticking probability dependence on dust temperature is taken according to \citet{He2016}. 

In shocks, the dust grain velocity may differ from that of the gas particles. Following \citet{Guillet2007}, we define effective temperature of accreted species:

\begin{equation}
T_{\rm{eff}} = T_{\rm{\alpha}} + \frac{\pi m_{\rm{j}}}{8k_{\rm{B}}} ({\bf{u}}_{\rm{g}} - {\bf{u}}_{\rm{\alpha}})^2,
\label{eff_temp_gas_dust}
\end{equation}

\noindent
where ${\bf{u}}_{\rm{g}}$ is the velocity vector of dust grains, ${\bf{u}}_{\rm{\alpha}}$ is the flow velocity vector of a gas component $\alpha$. The parameter $({\bf{u}}_{\rm{g}} - {\bf{u}}_{\rm{\alpha}})^2$ is averaged over grain charge distribution.

\subsubsection{Thermal desorption}
Several desorption mechanisms are included in the model. We introduce a grain coverage factor $\xi$ to take into account the fact that desorption occurs from the top $N_{\rm{act}}$ active monolayers only:

\begin{equation}
\xi = {\rm{min}} \left( \frac{4 \pi a^2 N_{\rm{act}} N_{\rm{s}}n_{\rm{g}}}{n_{\rm{ice}}}, 1 \right),
\label{eq_desorp_factor}
\end{equation}

\noindent
where $n_{\rm{ice}}$ is the total number density of adsorbed species, $n_{\rm{ice}} = \sum_{\rm{j}} n^{\rm{s}}_{\rm{j}}$, $n^{\rm{s}}_{\rm{j}}$ is the number density of adsorbed species of type $j$, $N_{\rm{s}}$ is the number of adsorption sites on dust grain per unit grain area. The number density of adsorbed species $j$ that are located at the grain surface is $\xi n^{\rm{s}}_{\rm{j}}$ under an assumption that adsorbed species are well-mixed in icy mantles of grains.
 
The rate of thermal desorption $k_{\rm{thd, j}}$ is given by \citep{Hasegawa1992}:

\begin{equation}
k_{\rm{thd, j}} = \nu_{\rm{s, j}} \, {\rm{exp}} \left( -E_{\rm{D, j}}/T_{\rm{d}} \right)  \ \left[{\rm{s^{-1}}}\right],
\label{therm_desorp_rate}
\end{equation}

\noindent
where $T_{\rm{d}}$ is the temperature of dust grains, $E_{\rm{D, j}}$ is the desorption or binding energy of species $j$ (in K), and $\nu_{\rm{s, j}}$ is the characteristic vibration frequency of the adsorbed species given by

\begin{equation}
\displaystyle
\nu_{\rm{s, j}} = \sqrt{\frac{2 N_{\rm{s}} k_{\rm{B}} E_{\rm{D, j}}}{\pi^2 m_{\rm{j}}}}.
\end{equation}

\noindent
The binding energies for chemical species are taken from \citet{Penteado2017}. In calculations of thermal evaporation rates of chemical species, we allow $E_{\rm{D, j}}$ values to be no greater than that of H$_2$O \citep{Collings2004}.

\subsubsection{Direct cosmic ray desorption}
We use the method by \citet{Roberts2007} to calculate the cosmic ray induced desorption rates. This method is different from that adopted in most of the models, i.e. the method by \citet{Hasegawa1993a}: the rate is calculated by considering the number of molecules capable of being desorbed per cosmic ray impact. The cosmic ray induced desorption rate is given by

\begin{equation}
k_{\rm{crd, j}} = \frac{\zeta}{\zeta_0} \frac{F_{\rm{CR}} \phi}{4 N_{\rm{act}} N_{\rm{s}}} \ \left[{\rm{s^{-1}}}\right],
\end{equation}

\noindent
where $\phi$ is the efficiency parameter such that $\phi \, n_{\rm{j}}^{\rm{s}}/n_{\rm{ice}}$ is the number of species of type $j$ released per cosmic ray impact, $F_{\rm{CR}}$ is the flux of iron nucleus component of cosmic rays, as the iron nuclei are expected to be the most effective in heating the grains, here we take $\phi = 10^5$ and $F_{\rm{CR}} = 2 \times 10^{-3}$~cm$^{-2}$~s$^{-1}$ \citep{Roberts2007}. Only volatile species such as CO, N$_2$, O$_2$, CH$_4$ are expected to be desorbed.

\subsubsection{Photo-desorption}
The photo-desorption is dependent on the composition of the ice and the spectrum of the UV radiation field \citep{Fayolle2011,Bertin2013}. As by \citet{Ruaud2016}, we use a simplistic approach and consider a single photo-desorption yield $Y$ for all species -- we take $Y = 10^{-4}$ species per photon. The rate of photo-desorption by interstellar UV photons is given by

\begin{equation}
k_{\rm{phd, j}} = G_0 \, {\rm{exp}}(-2A_{\rm{V}}) \frac{F_{\rm{UV}} Y}{4 N_{\rm{act}} N_{\rm{s}}} \ \left[{\rm{s^{-1}}}\right],
\label{photo_desorp_rate}
\end{equation}

\noindent
where $F_{\rm{UV}}$ is the 'standard' unshielded flux of the interstellar UV radiation field, we set $F_{\rm{UV}}$ = $2\times 10^8$~cm$^{-2}$~s$^{-1}$ \citep{Draine1978}. 

\subsubsection{Cosmic ray induced UV photo-desorption}
This photo-desorption process is identical to the photo-desorption considered above, except for the source of UV photons. The interaction of cosmic ray particles with the molecular gas leads to the excitation of upper electronic states of molecular hydrogen, followed by far-ultraviolet emission \citep{Prasad1983}. The rate of photo-desorption is given by

\begin{equation}
k_{\rm{phd-CR, j}} = \frac{\zeta}{\zeta_0} \frac{2n_{\rm{H_2}}}{n_{\rm{H, tot}}} \frac{F_{\rm{UV-CR}} Y}{4 N_{\rm{act}} N_{\rm{s}}} \ \left[{\rm{s^{-1}}}\right],
\label{CRphoto_desorp_rate}
\end{equation}

\noindent
where $F_{\rm{UV-CR}}$ is the 'standard' cosmic ray induced UV flux, here we take $F_{\rm{UV-CR}} = 3 \times 10^3$~cm$^{-2}$~s$^{-1}$ based on the results by \citet{Cecchi-Pestellini1992}.

\subsection{Grain surface chemistry}
In dense and shielded regions, icy mantles form around dust grain cores via accretion of gas-phase species, and chemical reactions on the surface of grains take place. We consider a two-phase chemical model (the gas and grain surface) in which there is no distinction between the outermost and inner monolayers of ice mantles of grains. 

\subsubsection{Two-body surface reactions}
We consider Langmuir--Hinshelwood mechanism of grain surface reactions. In this mechanism, two species migrate on the surface and react with each other upon an encounter. The reaction rate coefficient $k_{\rm{s, jl}}$ between adsorbed species $j$ and $l$ can be expressed as \citep{Hasegawa1992}:

\begin{equation}
k_{\rm{s, jl}} = \kappa_{\rm{jl}} \left( \frac{1}{t_{\rm{diff, j}}} + \frac{1}{t_{\rm{diff, l}}} \right) \frac{1}{n_{\rm{g}}} \ \left[{\rm{cm^{3} s^{-1}}}\right],
\label{surface_reaction_rate_coeff}
\end{equation}

\noindent
where $\kappa_{\rm{jl}}$ is the probability that the reaction occurs, $t_{\rm{diff, j}}$ is the diffusion time required for an adsorbed specimen $j$ to swap over a number of sites equivalent to the surface of one grain. The diffusion can be thermal when species migrate from one site to another one by thermal hopping, or non-thermal when species cross through the potential barrier by quantum tunnelling. The time of thermal hopping for species $j$ from one surface site to an adjacent site is \citep{Hasegawa1992}:

\begin{equation}
t_{\rm{hop, j}} = \nu_{\rm{s, j}}^{-1} {\rm{exp}} \left( E_{\rm{b,j}}/T_{\rm{d}} \right),
\end{equation} 

\noindent
the height $E_{\rm{b,j}}$ of the barrier against diffusion is assumed to be $0.35E_{\rm{D,j}}$ for all species \citep{Karssemeijer2014}. The time-scale $t_{\rm{q,j}}$ for species $j$ to migrate via quantum tunnelling through a rectangular barrier of thickness $a_{\rm{s}}$ is:

\begin{equation}
t_{\rm{q, j}} = \nu_{\rm{s, j}}^{-1} {\rm{exp}} \left[ 2(a_{\rm{s}}/ \hbar) \left( 2 m_{\rm{j}} k_{\rm{B}} E_{\rm{b,j}} \right)^{1/2} \right].
\end{equation} 

\noindent
The diffusion time $t_{\rm{diff, j}}$ is given by

\begin{equation}
t_{\rm{diff, j}} = 4 \pi a^2 N_{\rm{s}} t_{\rm{hop, j}}, 
\end{equation}

\noindent
and analogous expression with $t_{\rm{q, j}}$. We allow H and H$_2$ to migrate through thermal hopping or quantum tunnelling depending on which is faster. For other species, diffusion only via thermal hopping is considered.

For reactions without activation barrier, $\kappa_{\rm{jl}}$ is considered equal to 1 \citep{Hasegawa1992}. For reactions with activation barrier $E_{\rm{A, jl}}$, the 'reaction--diffusion' competition mechanism is employed and the probability $\kappa_{\rm{jl}}$ is given by \citep{Chang2007, Ruaud2016}:

\begin{equation}
\displaystyle
\kappa_{\rm{jl}} = \frac{\nu_{\rm{s,jl}} \widetilde{\kappa}_{\rm{{jl}}}}{\nu_{\rm{s,jl}} \widetilde{\kappa}_{\rm{{jl}}} + 1/t_{\rm{hop, j}} + 1/t_{\rm{hop, l}} + \displaystyle\sum k_{\rm{d, j}} + \displaystyle\sum k_{\rm{d, l}}},
\label{k_react_diff_comp}
\end{equation}

\noindent
where $k_{\rm{d, j}}$ is the rate coefficient of a desorption process, $\nu_{\rm{s,jl}}$ is taken to be equal to the higher value of characteristic vibration frequencies of two reactants $j$ and $l$. A sum over all desorption mechanisms must be taken, but here we take into account only thermal desorption. The parameter $\widetilde{\kappa}_{\rm{{jl}}}$ can be expressed as:

\begin{equation}
\widetilde{\kappa}_{\rm{{jl}}} = \exp \left( -E_{\rm{A, jl}}/T_{\rm{d}} \right),
\label{k_act_barr_classic}
\end{equation}

\noindent
or as the quantum mechanical probability for tunnelling through a rectangular barrier of thickness $a_{\rm{s}}$: 

\begin{equation}
\widetilde{\kappa}_{\rm{jl}} = \exp \left[ -2(a_{\rm{s}}/\hbar)(2 m_{\rm{jl}} k_{\rm{B}}E_{\rm{A, jl}})^{1/2} \right],
\label{k_act_barr_qtunn}
\end{equation}

\noindent
we use a value of 1.5~\AA\, for a barrier thickness \citep{Cuppen2017}. In calculating reaction rates we adopt the highest value of the equations (\ref{k_act_barr_classic}) and (\ref{k_act_barr_qtunn}) for any given dust temperature.

The grain surface network and reaction activation barriers are taken from the NAUTILUS code files \citep{Ruaud2016}. The rates of homogeneous reactions (that involve the same species) must be multiplied by the factor 0.5. Equal branching fractions are assumed for the reactions with multiple channels and without activation barrier. For reactions with activation barrier, the competition between different channels is considered. 

\subsubsection{Photo-dissociation and photo-ionization}
Photo-dissociation and photo-ionization of adsorbed species are included. Rate coefficients for these reactions are calculated according to the same equations as for gas-phase chemistry. We assume that ion and electron produced during the photo-ionization recombine instantaneously to form products on the surface with branching ratios based on the relevant gas-phase dissociative recombination reactions \citep{Ruffle2001}. The reaction parameters according to the UDfA chemical network are used.

\subsubsection{Reactive desorption}
For visual extinctions much higher than 1, reactive (chemical) desorption is expected to play important role in production of gaseous species such as CH$_3$OH and H$_2$O$_2$ \citep{Garrod2007,Du2012}. The efficiency of the reactive desorption on water ices is poorly constrained \citep{Minissale2016}. In our model, for each surface reaction that leads to a single product, a proportion $f$ of the product species is released into the gas phase, whilst the rest, $1 - f$, remains as a surface-bound product. The fraction $f$ is assumed to be equal to $f = 0.01$ for all qualifying reactions \citep{Chuang2018}.

\subsubsection{Encounter desorption mechanism for H$_2$ molecules}
\citet{Hincelin2015} proposed the 'encounter desorption' mechanism for molecular hydrogen on grain surface. An H$_2$ molecule can, while diffusing on the surface, find itself on another hydrogen molecule. The H$_2$--H$_2$ bond is far weaker than the H$_2$--water bond, that leads to efficient desorption.

The parameters of grain surface chemistry are given in the Table \ref{table:chemistryparam}.

\begin{table}
\caption{Parameters of grain surface chemistry.}
\begin{tabular}{ll}
\hline
Parameter & Value \\
\hline
Surface site density, $N_{\rm{s}}$ & $10^{15}$~cm$^{-2}$ \\
Diffusion to desorption energy ratio & 0.35 \\
Desorption energy of H$_2$ on H$_2$ substrate & 23~K $^{a}$\\
Photo-desorption yield, $Y$ & $10^{-4}$ \\
Efficiency of the reactive desorption, $f$ & 0.01 \\
Thermal hopping & On \\
Quantum tunnelling through diffusion barriers & On \\
Barrier thickness for diffusion & 1.5~\AA\ \\
Quantum tunnelling through reaction barriers & On \\
Barrier thickness for reaction & 1.5~\AA\ \\
Number of active monolayers, $N_{\rm{act}}$ & 2 \\ 
\hline
\end{tabular}

\small
\medskip
$^{a}${\citet{Cuppen2007}}
\label{table:chemistryparam} 
\end{table}

\subsection{Kinetic equations}
We use 'rate equations' method to study the evolution of chemical species in the gas and on the grain surface. The concentration of each specimen is obtained by solving differential equations. This approach is valid only if the average number of reactive species on the surface of one grain is large \citep{Cuppen2013}.

The rate of formation of species $j$ in a unit volume of gas is:

\begin{equation}
\begin{array}{l}
\displaystyle
\mathcal{N}_{\rm{j}} = \sum k^{\rm{j}}_{\rm{lm}} n_{\rm{l}} n_{\rm{m}} - n_{\rm{j}} \sum k_{\rm{jl}} n_{\rm{l}} + \sum k^{\rm{j}}_{\rm{l}} n_{\rm{l}} - n_{\rm{j}} \sum k_{\rm{j}} + \\ [15pt]
\displaystyle
+ \sum k^{\rm{j}}_{\rm{rd, lm}} n_{\rm{l}}^{\rm{s}} n_{\rm{m}}^{\rm{s}} + n_{\rm{j}}^{\rm{s}} \xi \sum k_{\rm{d, j}} + n_{\rm{j}}^{\rm{s}} \xi k_{\rm{sp, j}} - k_{\rm{a, j}} n_{\rm{j}}, \\ [15pt]
\displaystyle
\mathcal{N}_{\rm{j}}^{\rm{s}} = \sum k^{\rm{j}}_{\rm{s, lm}} n_{\rm{l}}^{\rm{s}} n_{\rm{m}}^{\rm{s}} - n_{\rm{j}}^{\rm{s}} \sum k_{\rm{s,jl}} n_{\rm{l}}^{\rm{s}} + \sum k_{\rm{s,l}}^{\rm{j}} n_{\rm{l}} - n_{\rm{j}}^{\rm{s}} \sum k_{\rm{s,j}} \\ [15pt]
\displaystyle
 - n_{\rm{j}}^{\rm{s}} \sum k_{\rm{rd, jl}}  n_{\rm{l}}^{\rm{s}} - n_{\rm{j}}^{\rm{s}} \xi \sum k_{\rm{d, j}} - n_{\rm{j}}^{\rm{s}} \xi k_{\rm{sp, j}} + k_{\rm{a, j}} n_{\rm{j}},
\end{array}
\label{chem_kin_eqs}
\end{equation}

\noindent
where $k^{\rm{j}}_{\rm{lm}}$ denotes the rate coefficient for a gas-phase bimolecular reaction between species $l$ and $m$ having the specimen $j$ as a product, and the sum is over all such reactions, $k^{\rm{j}}_{\rm{l}}$ is the rate coefficient for a gas-phase unimolecular reaction with a product $j$; $k^{\rm{j}}_{\rm{s, lm}}$ and $k_{\rm{s,l}}^{\rm{j}}$ have the same meaning but for surface reactions; $k^{\rm{j}}_{\rm{rd, lm}}$ is the rate coefficient for surface reaction leading to desorption of species $j$, $k_{\rm{sp, j}}$ is the sputtering rate of ice mantle species (see section \ref{sect_sputtering}). We introduce a factor $\xi$ (equation (\ref{eq_desorp_factor})) in the desorption and sputtering terms to take into account the specimen escape from top ice layers.

\subsection{Dust charging}
It has been shown that grain charging (especially charging of small grains or PAH molecules) can affect the gas-phase chemistry in dense molecular clouds \citep{Dalgarno2006,Wakelam2008,Kochina2014}. Moreover, grain charging processes are critical for the study of dynamics and processing of dust grains in C-type shocks because they affect the coupling of grains to the ion fluid \citep{Guillet2007}. Hence, the dust charging and ionization balance of the gas must be calculated jointly.

\subsubsection{Photoelectric emission}
The photoelectric emission rate and energy of ejected photoelectrons are calculated within the framework described by \citet{Weingartner2001b}. The photoelectric emission rates are calculated for the unshielded interstellar UV radiation field \citep{Draine1978}, and for the unshielded interstellar radiation at optical wavelengths \citep{Mathis1983,Draine2011}. These rates are scaled for a given visual extinction. Note, that negatively charged small grains (PAH molecules) are effectively neutralized by visible photons \citep{Wakelam2008}.

The spectrum of cosmic ray induced UV radiation consists of many individual lines, photons are emitted in the energy range 7.1--14.6~eV \citep{Gredel1989}. In our estimates of photoelectric emission rates, we assume that the photon flux is independent on energy. The 'standard' photon flux is equal to $F_{\rm{UV-CR}}$. The photoelectric emission rates are scaled depending on the cosmic ray ionization rate.

\subsubsection{Collisional attachment of ions and electrons}
The accretion rate of particles $j$ with charge $q_{\rm{j}}$ on a grain having radius $a$ and charge $Ze$ is \citep{Draine1987}:

\begin{equation}
\displaystyle
J_{\rm{j}}(a,Z) = n_{\rm{j}} s_{\rm{j}} \left( \frac{8 k_{\rm{B}} T}{\pi m_{\rm{j}}}\right)^{1/2} \pi a^2 \widetilde{J}(a k_{\rm{B}}T/q_{\rm{j}}^2, Ze/q_{\rm{j}}) \ \left[{\rm{s^{-1}}}\right],
\label{accr_rate}
\end{equation}

\noindent
where $s_{\rm{j}}$ is the probability that specimen $j$ will transfer its charge if it reaches the surface of a grain, $T$ is the kinetic temperature of species $j$, $e$ is the elementary charge. Formulae for the reduced rate coefficient $\widetilde{J}$ can be found in \citet{Draine1987}. 

We assume that positive ion has a high probability of seizing an electron if it arrives at the surface of a grain, $s_{\rm{j}} = 1$ \citep{Weingartner2001b}. We use effective ion temperature defined by the equation (\ref{eff_temp_gas_dust}) in computing the arrival rate of ions to the grain surface. We consider caution--grain recombination channels following \citet{Wakelam2008}. For grain--electron collisions, we use the equation (\ref{accr_rate}) with electron temperature $T_{\rm{e}}$. The sticking probability for electrons $s_{\rm{e}}$ is taken equal to 0.5 for large grains \citep{Weingartner2001b}.

\subsubsection{Grain charge distribution}
As by \citet{Guillet2007}, our method for charge integration has two modes. Weakly charged grains ($\vert Z \vert \lesssim 10$) have their charge distribution. For highly charged grains, the average charge of the grains is calculated. Let $n_{\rm{g},Z}$ to be the number density of grains having the charge $Ze$. The rate of production of grains with charge $Ze$ in a unit volume of gas is:

\begin{equation}
\begin{array}{l}
\displaystyle
\mathcal{N}_{\rm{g},Z} = n_{\rm{g},Z+1} J_{\rm{e}}(a,Z+1) + n_{\rm{g},Z-1} \left[ J_{\rm{ph}}(a,Z-1) + \sum_{j} J_{\rm{j}}(a,Z-1) \right] - \\ [15pt]
\displaystyle
- n_{\rm{g},Z} \left[ J_{\rm{e}}(a,Z) + J_{\rm{ph}}(a,Z) + \sum_{j} J_{\rm{j}}(a,Z) \right],
\end{array}
\label{grain_distr_kin_eq}
\end{equation}
\noindent
where $J_{\rm{ph}}(a,Z)$ is the photoelectron emission rate for the dust grain having the radius $a$ and the charge $Ze$, the sum is over all positive ions for which ion--grain recombination reactions are considered.

If the average charge of grains becomes high, $\vert \langle Z \rangle \vert > Z_{\rm{max}}$, the charge distribution integration is stopped and is replaced by the integration of the average charge $\langle Z \rangle$. In this case, the rate of grain charge change is:

\begin{equation}
\frac{{\rm{d}} \langle Z \rangle}{{\rm{d}}t} = J_{\rm{ph}}(a,\langle Z \rangle) + \sum_{\rm{j}} J_{\rm{j}}(a,\langle Z \rangle) - J_{\rm{e}}(a,\langle Z \rangle).
\label{av_grain_charge_eq}
\end{equation} 

\noindent
The integration of charge distribution replaces the average charge integration if $\vert \langle Z \rangle \vert < Z_{\rm{max}}-1$. In the calculations, $Z_{\rm{max}} = 16$.

\section{Calculation of ion and molecule level populations}
\label{app_levpop}
\subsection{Spectroscopic data and collisional rate coefficients}
The level populations of ions CI, CII, OI and molecules H$_2$, CO, H$_2$O are evaluated. We take into account five lower levels of OI -- $^3P_2$, $^3P_1$, $^3P_0$, $^1D_2$, and $^1S_0$; three levels of CI -- $^3P_0$, $^3P_1$, $^3P_2$; two levels of CII -- $^2P^{\circ}_{1/2}$ and $^2P^{\circ}_{3/2}$. The spectroscopic data for ions are taken from NIST Atomic Spectra Database\footnote{\url{https://www.nist.gov/pml/atomic-spectra-database}}. We take into account 150 rotational levels of H$_2$ molecule. The level energies of H$_2$ are taken from \citet{Dabrowski1984} and Einstein coefficients are taken from \citet{Wolniewicz1998}. We take into account 150 rotational levels of ortho-H$_2$O and 150 rotational levels of para-H$_2$O belonging to the ground and first vibrational excited states of the molecule (the energy of the highest level is about 4500~K). In the case of CO molecule, we take into account 41 rotational levels of the ground vibrational state. The spectroscopic data for CO and H$_2$O molecules are taken from the HITRAN database \citep{Rothman2013*,Gordon2017}.

Data on collisional rate coefficients used in the simulations are listed in the Table \ref{table:collcoeff}. The excitation of H$_2$ by thermal electrons is the dominant mechanism of electron cooling in shocked molecular gas. For the collisions of H$_2$ with thermal electrons, the rate coefficients are estimated using the data given by \cite{Gerjuoy1955,Ehrhardt1968,England1988,Yoon2008} and recipes given by \citet{Faure2008}. In our model, ortho-/para- states of molecular hydrogen may change in reactive collisions with hydrogen atoms that is relevant only for the hot postshock gas. The rate coefficients for CII--He collisions are taken equal to those for CII--H collisions multiplied by 0.38 \citep{Draine2011}. Most of the collisional data used in calculations are available in the BASECOL database\footnote{\url{http://basecol.obspm.fr}} \citep{Dubernet2013}.  

At kinetic temperatures above the maximal temperature for which the collisional coefficients are given, their values are assumed to remain constant \citep{Flower2012}. At temperatures lower than the minimal temperature, collisional coefficients are assumed to be proportional to gas kinetic temperature.

\begin{table}
\caption{Data on collisional rate coefficients.}
\begin{tabular}{lll}
\hline
Specimen & Collisional partner & Reference \\
\hline
OI & H$_2$ & \citet{Jaquet1992} \\
 & & \citet{Glover2007} \\
 & He & \citet{Monteiro1987} \\
 & H & \citet{Krems2006} \\
 & & \citet{Abrahamsson2007} \\
 & e$^{-}$ & \citet{Berrington1981} \\
 & & \citet{Pequignot1990} \\
 & & \citet{Bell1998} \\ [5pt]
CI & H$_2$ & \citet{Schroder1991} \\
 & He & \citet{Staemmler1991} \\
 & H & \citet{Abrahamsson2007} \\
 & e$^{-}$ & \citet{Johnson1987} \\ [5pt]
CII & H$_2$ & \citet{Flower1977} \\
 & & \citet{Draine2011} \\
 & H & \citet{Barinovs2005} \\
 & e$^{-}$ & \citet{Tayal2008} \\
 & & \citet{Draine2011} \\ [5pt]
H$_2$ & H$_2$ & \citet{Flower1998b} \\
 & & \citet{Flower1999} \\
 & He & \citet{Flower1998a} \\
 & H & \citet{Lique2015} \\
 & & \citet{Wrathmall2007} \\
 & & \citet{Martin1995} \\ 
 & e$^{-}$ & \cite{Gerjuoy1955} \\
 & & \cite{Ehrhardt1968} \\
 & & \cite{England1988} \\ 
 & & \citet{Yoon2008} \\ [5pt]
CO & H$_2$ & \citet{Yang2010} \\
 & He & \citet{Cecchi-Pestellini2002} \\
 & H & \citet{Walker2015} \\ [5pt]
H$_2$O & H$_2$ & \citet{Faure2007} \\
 & & \citet{Faure2008} \\
 & He & \citet{Green1993} \\
 & & \citet{Nesterenok2013} \\
 & H & \citet{Daniel2015} \\
 & e$^{-}$ & \citet{Faure2008} \\
\hline
\end{tabular}
\label{table:collcoeff}
\end{table}

\subsection{System of master equations for level population densities of ions and molecules}
The assumption of statistical equilibrium on level population densities may not be valid in dynamically active regions, where physical parameters vary rapidly \citep{Flower2009}. The population densities of energy levels of ions and molecules are computed in parallel with dynamical and chemical rate equations. The system of differential equations describing the evolution of the level population densities of specimen $j$ is

\begin{equation}
\displaystyle
\frac{\rm{d}}{{\rm{d}}z} \left( \chi_{\rm{j,m}} u_{\rm{\alpha}} \right) = \sum_{l=1, \, l \ne m}^M \left( R_{\rm{lm}} + C_{\rm{lm}} \right) \chi_{\rm{j,l}} - \chi_{\rm{j,m}} \sum_{l=1, \, l \ne m}^M \left( R_{\rm{ml}} + C_{\rm{ml}} \right) +  \bar{\chi}_{\rm{j,m}} \mathcal{N}_{\rm{j}}, \quad m=1,...,M,
\label{level_pop_eqn}
\end{equation}

\noindent
where $\chi_{\rm{j,m}}$ is the population density (in cm$^{-3}$) of level $m$, $M$ is the total number of levels, $R_{\rm{ml}}$ is the rate coefficient for the transition from level $m$ to level $l$ through radiative processes, and $C_{\rm{ml}}$ is the rate coefficient of collisional processes, $\mathcal{N}_{\rm{j}}$ is the production (or destruction) rate of specimen $j$ in chemical reactions, $\bar{\chi}_{\rm{j,m}}$ is the population distribution of energy levels of newly formed species, $u_{\rm{\alpha}}$ is the flow velocity of neutral fluid if the specimen $j$ is neutral or ion flow velocity for ion species (e.g. CII). For all chemical reactions, we assume that newly formed molecules or ions have population distribution of energy levels proportional to the current population density distribution $\chi_{\rm{j,m}}$. 

The system of equations (\ref{level_pop_eqn}) must be completed by radiation transfer equation in molecular (atomic) lines. We calculate radiation intensity in molecular lines using the large velocity gradient (or Sobolev) approximation \citep{Sobolev1957,Hummer1985}. Here, we assume that in the shock region where velocity gradient is high, the gas temperature is much higher then the dust temperature. Consequently, we disregard the dust emission in calculations of radiation intensity. Consider an one-dimensional flat gas--dust cloud with a constant gas velocity gradient. Let $\nu_{\rm{lm}}$ to be the average line frequency and $\Delta\nu_{\rm{lm}}$ to be the line profile width that is determined by the spread in thermal velocities of molecules and micro-turbulence, 

\begin{equation}
\Delta\nu_{\rm{lm}}=\nu_{\rm{lm}}\frac{v_{\rm{D}}}{c}, \quad v_{\rm{D}}^{2} = v_{\rm{th}}^2 + v_{\rm{turb}}^2,
\nonumber
\end{equation}

\noindent
where $v_{\rm{th}}$ is the most probable value of the thermal speed of species in question, and $v_{\rm{turb}}$ is the characteristic micro-turbulence speed in the cloud. The length of the region where the flow velocity changes by the value corresponding to the line width is:

\begin{equation}
\Delta z_{\rm{D}} = \frac{v_{\rm{D}}}{{\rm{d}}u_{\rm{\alpha}} / {\rm{d}}z}, 
\end{equation}

\noindent
where $u_{\rm{\alpha}}$ is the flow velocity (of neutrals or ions depending on specimen ionic state). The mean intensity $J_{\rm{lm}}(z)$ in molecular line is given by \citep{Hummer1985}:

\begin{equation}
J_{\rm{lm}}(z) = S_{\rm{L}}(z) \left[1 - 2\mathscr{P} \left( \gamma_{\rm{L}}, \gamma_{\rm{C}} \right) \right],
\label{line_int_lvg}
\end{equation}

\noindent
where $S_{\rm{L}}$ is the line source function, the parameter $\mathscr{P}$ is one-sided loss probability for photons created by line processes, $\gamma_{\rm{L}}$ and $\gamma_{\rm{C}}$ are line parameters,

\begin{equation}
\frac{1}{\gamma_{\rm{L}}} = k_{\rm{L}} \Delta z_{\rm{D}}, \quad \frac{1}{\gamma_{\rm{C}}} = k_{\rm{C}} \Delta z_{\rm{D}},
\end{equation}

\noindent
where $k_{\rm{C}}$ is the absorption coefficient of the dust at the line frequency, $k_{\rm{L}} \phi(x)$ is the line opacity coefficient, $\phi(x)= \rm{exp}(-x^2)/\pi^{1/2}$, $x = (\nu - \nu_{\rm{lm}})/\Delta\nu_{\rm{lm}}$. The details of the calculations of the loss probability function see in \citet{Nesterenok2016}.

\section{Heating and cooling processes of gas and dust}
\label{app_heatcool}
\subsection{Gas cooling by atomic and molecular line emission}
\label{sss_gas_cool}
The gas loses thermal energy through the emission in atomic and molecular lines. The amount of thermal energy that is transferred from the gas to the excitation of energy levels of a specimen $j$ is:

\begin{equation}
\begin{array}{l}
\displaystyle
\mathcal{G}_{\rm{j, rad}} = \sum_{\rm{l > m}} h \nu_{\rm{lm}} \left( C_{\rm{lm}} \chi_{\rm{j,l}} - C_{\rm{ml}} \chi_{\rm{j,m}} \right),
\end{array}
\label{line_cooling_rate}
\end{equation}

\noindent
where summation is over all collisional transitions of a specimen $j$, $\mathcal{G}_{\rm{j, rad}} < 0$ for cooling process. Collisions of a specimen with neutral partners H$_2$, He and H cool (heat) neutral gas component, while collisions with electrons cool (heat) the electron gas.

\subsection{Heating by cosmic rays and photoelectric heating}
Cosmic ray particles ionize atoms and molecules of the medium, as a result energetic electrons are produced. Energy degradation of fast charged particles in materials is characterized by a mean energy per ion pair $W$, which is the initial energy of the particle divided by the number of ionizations. For fast electrons propagating in the interstellar molecular gas $W \simeq 35$~eV \citep{Dalgarno1999}. The heating rate of the gas can be estimated:

\begin{equation}
\mathcal{G}_{\rm{n,CR}} = n_{\rm{H,tot}} \zeta \eta W,
\end{equation}

\noindent
where $\eta$ is the heating efficiency defined as the fraction of the primary particle energy that is converted into heat. We use data on heating efficiency presented by \citet{Dalgarno1999} -- $\eta \simeq 0.05$ at high energy limit and low fractional ionizations, $x_{\rm{e}} < 10^{-4}$. The rotational excitation of H$_2$ molecules, Coulomb losses and momentum transfer with neutral particles are considered as a heat source by \citet{Dalgarno1999}. We assume that all this heat goes to neutral fluid.

The heating of the gas by photoelectric emission from dust grains (due to absorption of interstellar and cosmic ray induced UV radiation fields) is taken into account \citep{Weingartner2001b}. We assume that all energy of photoelectrons goes to neutral gas.

\subsection{Chemical heating}
The heat released in exothermic gas-phase chemical reactions is important heating mechanism in interstellar clouds at high visual extinctions \citep{LeBourlot1993,Glassgold2012}. The main contribution to the chemical heating comes from the dissociative recombination of the abundant molecular ions, some ion--neutral reactions, and H$_2$ formation on grains. We assume that 30 per cent of the chemical energy released in dissociative recombination reactions goes to the kinetic energy of the reaction products \citep{Ojekull2004,Hamberg2014}. We use recipes given by \citet{Draine1986} in the calculations of momentum and heat source terms associated with chemical reactions.

\subsection{Ion--electron scattering}
Energy transfer between ions and electrons is taken into account. The expression for energy exchange rate is given by \citet{Spitzer1956,Draine1980}.

\subsection{Thermal balance of dust}
The dust temperature is a key parameter in the grain-surface chemistry simulations. We take into account dust cooling by self-emission, dust heating by gas--dust collisions (see section \ref{sect_grain_dynam}), and by interstellar radiation field. The heating rate of a dust particle by interstellar radiation field is given by the equation:

\begin{equation}
\displaystyle
\mathcal{G}_{\rm{d,IS}} = 4\pi \int\limits_0^{\infty} {\rm{d}}\nu \, C_{\rm{abs}}(\nu) I_{\rm{IS}}(\nu) {\rm{exp}} \left[-\tau_{\rm{ext}}(\nu)\right],
\label{grain_heat_is}
\end{equation}  

\noindent
where $C_{\rm{abs}}(\nu)$ is the absorption cross section of a dust grain, $I_{\rm{IS}}(\nu)$ is the unshielded intensity of the interstellar radiation field, $\tau_{\rm{ext}}(\nu)$ is the optical depth to the cloud boundary that corresponds to the $A_{\rm{V}}$ in question. We use the dust extinction law that is representative for cold dark clouds and corresponds to the ratio of visual extinction to reddening $R_V = 5.5$ \citep{Weingartner2001,Draine2003}. The approximation formulae for the intensity of local interstellar starlight background is taken from \citet{Draine1978,Mathis1983,Draine2011}, for interstellar dust emission -- from \citet{Hocuk2017}. 

The grain cooling rate via self-emission $\mathcal{G}_{\rm{d,rad}}$ is:

\begin{equation}
\mathcal{G}_{\rm{d,rad}} = -4\pi \int\limits_0^{\infty} {\rm{d}} \nu \, C_{\rm{abs}}(\nu) B(\nu, T_{\rm{d}}),
\end{equation}

\noindent
where $B(\nu, T_{\rm{d}})$ is the Planck function. The evolution of dust temperature is described by the equations: 

\begin{equation}
\begin{array}{l}
\displaystyle
C_{\rm{d}} \frac{\rm{d}T_{\rm{d}}}{{\rm{d}}t} = \sum_k \mathcal{G}_{\rm{d,k}}, \\[15pt]
\displaystyle
u_{z{\rm{g}}} \frac{{\rm{d}}T_{\rm{d}}}{{\rm{d}}z} = \frac{{\rm{d}}T_{\rm{d}}}{{\rm{d}}t},  
\end{array}
\label{eq_dust_temp_evol}
\end{equation}

\noindent
where $C_{\rm{d}}$ is the heat capacity of the dust grain, $\mathcal{G}_{\rm{d,k}}$ is the dust grain heating or cooling rate by process $k$, $u_{z\rm{g}}$ is the average velocity of grain particles in $z$ direction. The heat capacity of dust grain is calculated based on Debye model \citep{Draine2001}.

\subsection{Chemical and thermal evolution of the static cloud}
In the case of a static cloud, the differential equations for specimen concentrations are:

\begin{equation}
\frac{{\rm{d}} n_{\rm{j}}}{{\rm{d}}t} = \mathcal{N}_{\rm{j}}, \quad \frac{{\rm{d}} n_{\rm{j}}^{\rm{s}}}{{\rm{d}}t} = \mathcal{N}_{\rm{j}}^{\rm{s}}, 
\end{equation}

\noindent
where $\mathcal{N}_{\rm{j}}$ and $\mathcal{N}_{\rm{j}}^{\rm{s}}$ are given by equations (\ref{chem_kin_eqs}). The equations for the time evolution of the grain charge distribution are:

\begin{equation}
\frac{{\rm{d}} n_{\rm{g},Z}}{{\rm{d}}t} = \mathcal{N}_{{\rm{g}},Z},
\end{equation}
\noindent
where $\mathcal{N}_{{\rm{g}},Z}$ are given by the equation (\ref{grain_distr_kin_eq}). Similarly, the equations for the time evolution of population number densities of energy levels of chemical species can be computed. Kinetic temperatures of gas components are obtained by solving equations:

\begin{equation}
\displaystyle
\frac{3}{2} k_{\rm{B}} \frac{\rm{d}}{{\rm{d}}t} \left(\frac{\rho_{\rm{\alpha}}}{\mu_{\rm{\alpha}}} T_{\rm{\alpha}} \right) = \sum_{k} \mathcal{G}_{\rm{\alpha,k}},
\end{equation}

\noindent
where $\alpha$ stands for neutral gas component, ions or electrons, $\mathcal{G}_{\rm{\alpha,k}}$ is gas component heating (or cooling) rate in process $k$.

\section{Shock model}
\label{app_shock}
\subsection{Magnetohydrodynamic equations}
A steady, one-dimensional flow along the $z$ direction is considered. The magnetic field is adopted to be in the $y$ direction -- transverse to the flow velocity. The shock velocity is equal to $u_{\rm{s}}$ and the magnetic field strength in the preshock gas is $B_0$. The flow components considered are neutrals, ions, and electrons, denoted by subscripts $n$, $i$, and $e$, respectively. Here we adopt that electrons and ions move at the same velocity, but kinetic temperatures of ions and electrons may differ \citep{Draine1980}. The assumption is made that magnetic field lines are frozen into the ion fluid, and the magnetic field in the postshock gas satisfies the equation \citep{Draine1986}:

\begin{equation}
B = B_0 \left( u_{\rm{s}}/u_{\rm{i}} \right).
\end{equation}  

The fluid equations that express conservation of particle number and mass densities are \citep{Draine1983}:

\begin{equation}
\begin{array}{l}
\displaystyle
\frac{\rm{d}}{{\rm{d}}z} \left( \frac{\rho_{\rm{n}}}{\mu_{\rm{n}}} u_{\rm{n}} \right) = \mathcal{N}_{\rm{n}} \\ [15pt]
\displaystyle
\frac{\rm{d}}{{\rm{d}}z} \left( \frac{\rho_{\rm{i}}}{\mu_{\rm{i}}} u_{\rm{i}} \right) = \mathcal{N}_{\rm{i}} \\ [15pt]
\displaystyle
\frac{\rm{d}}{{\rm{d}}z} \left( \rho_{\rm{n}} u_{\rm{n}} \right) = \mathcal{S}_{\rm{n}} \\ [15pt]
\displaystyle
\frac{\rm{d}}{{\rm{d}}z} \left( \rho_{\rm{i}} u_{\rm{i}} \right) = \mathcal{S}_{\rm{i}},
\end{array}
\label{mhd_shock_eqs1}
\end{equation}

\noindent
where $\rho_{\rm{n}}$ and $\rho_{\rm{i}}$ are mass densities of neutral and ion fluids, respectively, $\mu_{\rm{n}}$ is the mean mass per neutral particle and $\mu_{\rm{i}}$ is the mean mass per ion particle. The source terms at the right-hand side of the equations are defined as follows: $\mathcal{N}_{\rm{n}}$ and $\mathcal{N}_{\rm{i}}$ are the numbers of neutral particles and ions, respectively, created per unit volume and time as a result of chemical reactions; $\mathcal{S}_{\rm{n}}$ and $\mathcal{S}_{\rm{i}}$ are the rates per volume at which neutral and ion-electron masses are created, respectively.

The equations that express conservation of momentum are \citep{Draine1983}: 

\begin{equation}
\begin{array}{l}
\displaystyle
\frac{\rm{d}}{{\rm{d}}z} \left( \rho_{\rm{n}} u_{\rm{n}}^{2} +  k_{\rm{B}} T_{\rm{n}} \frac{\rho_{\rm{n}}}{\mu_{\rm{n}}}  \right) = \mathcal{F}_{\rm{n}} \\ [15pt]
\displaystyle
\frac{\rm{d}}{{\rm{d}}z} \left( \rho_{\rm{i}} u_{\rm{i}}^{2} +  k_{\rm{B}} T_{\rm{i}} \frac{\rho_{\rm{i}}}{\mu_{\rm{i}}} + k_{\rm{B}} T_{\rm{e}} n_{\rm{e}} + \frac{B_{0}^2 u_{\rm{s}}^{2}}{8 \pi u_{\rm{i}}^2} \right) = -\mathcal{F}_{\rm{n}}, \\ [15pt]
\end{array}
\label{mhd_shock_eqs2}
\end{equation}

\noindent
where $T_{\rm{n}}$, $T_{\rm{i}}$, and $T_{\rm{e}}$ are temperatures of gas components, $\mathcal{F}_{\rm{n}}$ is the rate per volume at which momentum is transferred from the charged fluid to the neutral fluid as a result of elastic scattering and chemical reactions.

The equations that express conservation of energy are \citep{Draine1983}:
\begin{equation}
\begin{array}{l}
\displaystyle
\frac{\rm{d}}{{\rm{d}}z} \left( \frac{1}{2} \rho_{\rm{n}} u_{\rm{n}}^{3} + \frac{5}{2} k_{\rm{B}} T_{\rm{n}} \frac{\rho_{\rm{n}}}{\mu_{\rm{n}}} u_{\rm{n}} \right) = \mathcal{G}_{\rm{n}} + \mathcal{F}_{\rm{n}} u_{\rm{n}} - \frac{1}{2} \mathcal{S}_{\rm{n}} u_{\rm{n}}^{2} \\ [15pt]
\displaystyle
\frac{\rm{d}}{{\rm{d}}z} \left( \frac{1}{2} \rho_{\rm{i}} u_{\rm{i}}^{3} + \frac{5}{2} k_{\rm{B}} T_{\rm{i}} \frac{\rho_{\rm{i}}}{\mu_{\rm{i}}} u_{\rm{i}} + \frac{5}{2} k_{\rm{B}} T_{\rm{e}} n_{\rm{e}} u_{\rm{i}} + \frac{B_{0}^2 u_{\rm{s}}^{2}}{4 \pi u_{\rm{i}}} \right) = \mathcal{G}_{\rm{i}} + \mathcal{G}_{\rm{e}} - \mathcal{F}_{\rm{n}} u_{\rm{i}} - \frac{1}{2} \mathcal{S}_{\rm{i}} u_{\rm{i}}^{2} \\ [15pt]
\displaystyle
\frac{\rm{d}}{{\rm{d}} z} \left( \frac{3}{2} k_{\rm{B}} T_{\rm{i}} \frac{\rho_{\rm{i}}}{\mu_{\rm{i}}} u_{\rm{i}} - \frac{3}{2} k_{\rm{B}} T_{\rm{e}} n_{\rm{e}} u_{\rm{i}} \right) + \left( k_{\rm{B}} T_{\rm{i}} \frac{\rho_{\rm{i}}}{\mu_{\rm{i}}} - k_{\rm{B}} T_{\rm{e}} n_{\rm{e}} \right) \frac{ {\rm{d}}u_{\rm{i}} }{{\rm{d}}z} = \mathcal{G}_{\rm{i}} - \mathcal{G}_{\rm{e}},
\end{array}
\label{mhd_shock_eqs3}
\end{equation}

\noindent
where $\mathcal{G}_{\rm{n}}$, $\mathcal{G}_{\rm{i}}$ and $\mathcal{G}_{\rm{e}}$ are rates, per volume, at which thermal energy is added to the neutral, ion and electron fluids, respectively. The change in internal energy of particles is not written in the left side of equations explicitly, but is taken into account in the heat source terms (see section \ref{sss_gas_cool}). The derivatives of the parameters $u_{\rm{n}}$, $u_{\rm{i}}$, $T_{\rm{n}}$, $T_{\rm{i}}$, and $T_{\rm{e}}$ are deduced from the system of differential equations given above \citep{Roberge1990}. 

The abundances of all atomic and molecular species are to be calculated together with magnetohydrodynamic equations,

\begin{equation}
\frac{\rm{d}}{{\rm{d}}z} \left( n_{\rm{j}} u_{\rm{\alpha}} \right) = \mathcal{N}_{\rm{j}},
\end{equation}

\noindent
where $u_{\rm{\alpha}}$ is the flow velocity of neutrals, ions, or average velocity of grains in $z$ direction depending on the specimen type; $\mathcal{N}_{\rm{j}}$ is the rate at which number density of species $j$ changes through chemical reactions, $\mathcal{N}_{\rm{j}}$ is given by equations (\ref{chem_kin_eqs}).

\subsection{Grain dynamics}
\label{sect_grain_dynam}
In the approach, adopted here, dust grains are treated as test particles. Charged grains enter the magnetic precursor and begin gyrating around magnetic field lines. \citet{Guillet2007} showed that gyration phase is short in dense gas and ends before a significant part of ice mantle has been eroded. Here, we ignore the gyration phase of charged grains. As gyration of charged grains has disappeared, grains reach a velocity that depends only on grain properties and local physical conditions in the medium. The general expression for fluid velocity of grains is given by \citet{Draine1980,Chapman2006}. 

The evolution of the grain charge distribution is described by the system of equations:

\begin{equation}
\frac{\rm{d}}{{\rm{d}}z} \left( n_{{\rm{g}},Z} \, u_{z{\rm{g}},Z} \right) = \mathcal{N}_{{\rm{g}},Z},
\end{equation}

\noindent
where $u_{z{\rm{g}},Z}$ is the flow velocity of grains having the charge $Ze$ in the $z$ direction, source term $\mathcal{N}_{{\rm{g}},Z}$ is given by the equation (\ref{grain_distr_kin_eq}). The concentration and charge of grains in the 'average charge' mode are governed by the equations:

\begin{equation}
\begin{array}{l}
\displaystyle
\frac{\rm{d}}{{\rm{d}}z} \left( n_{\rm{g}} u_{z{\rm{g}},\langle Z \rangle} \right) = 0 \\ [15pt]
\displaystyle
\frac{{\rm{d}} \langle Z \rangle}{{\rm{d}}z} = \frac{1}{u_{z{\rm{g}},\langle Z \rangle}} \frac{{\rm{d}} \langle Z \rangle}{{\rm{d}}t},
\end{array}
\end{equation}

\noindent
where the rate of charge change ${\rm{d}} \langle Z \rangle/{\rm{d}}t$ is given by the equation (\ref{av_grain_charge_eq}). 

In calculations, the rates of production and destruction of charged particles satisfy the charge conservation equation:

\begin{equation}
-\mathcal{N}_{\rm{e}} + \sum_i \mathcal{N}_{\rm{i}} + \sum_Z \mathcal{N}_{{\rm{g}},Z} = 0.
\end{equation}

\noindent
The large grains may not be coupled to the ion fluid and move somewhat faster. A local electric charge and electric field are created \citep{Guillet2007}. We neglect this electric field in our modelling. To satisfy charge neutrality of the medium, additional source of electrons is added for the electron number density:

\begin{equation}
\widetilde{\mathcal{N}_{\rm{e}}} = \sum_Z  \frac{\rm{d}}{{\rm{d}}z} \left[ Z n_{{\rm{g}},Z} \left( u_{\rm{i}} - u_{z{\rm{g}},Z} \right) \right],
\label{add_source_electrons}
\end{equation}

\noindent
where the sum is over the charge distribution of grains. For model parameters considered here, the relation usually holds $\vert \widetilde{\mathcal{N}_{\rm{e}}}\vert < 10^{-2} \vert\mathcal{N}_{\rm{e}}\vert$.

The rate of momentum transfer along the $z$ direction to the neutral gas through the friction between the gas and grains is \citep{Draine1980,Draine1986}:

\begin{equation}
\displaystyle
\mathcal{F}_{\rm{n, d}} = \sum_{Z} n_{{\rm{g}},Z} \, m_{\rm{g}} \left( u_{z{\rm{g}},Z} - u_{\rm{n}}\right) \frac{1}{\tau} = \sum_{Z} n_{{\rm{g}},Z} \, \sigma_{\rm{g}} \, \rho_{\rm{n}} \left( u_{z{\rm{g}},Z} - u_{\rm{n}} \right) \left[ \frac{128}{9 \pi} \frac{k_{\rm{B}} T_{\rm{n}}}{\mu_{\rm{n}}} + \left( {\bf{u}}_{{\rm{g}},Z} - {\bf{u}}_{\rm{n}} \right)^2 \right]^{1/2},
\end{equation}

\noindent
where $m_{\rm{g}}$ is the mass of one grain, $\tau$ -- viscous damping time. 

The rate at which the neutral gas is heated by collisions with dust particles is \citep{Draine1980}:

\begin{equation}
\mathcal{G}_{\rm{n, d}} = \mathcal{F}_{\rm{n, d}} \left( u_{\rm{i}} - u_{\rm{n}} \right) - \mathcal{L}_{\rm{d, n}},
\label{neutral_heat_grain_coll}
\end{equation}
\noindent
where $\mathcal{L}_{\rm{n,d}}$ is the dust heating rate due to grain--gas inelastic collisions,

\begin{equation}
\displaystyle
\mathcal{L}_{\rm{d,n}} = \sum_{Z} n_{{\rm{g}},Z} \sigma_{\rm{g}} \frac{\rho_{\rm{n}}}{\mu_{\rm{n}}} \left( \frac{8 k_{\rm{B}} T_{\rm{eff}}}{\pi \mu_{\rm{n}}} \right)^{1/2} \alpha_{\rm{d}}(T_{\rm{eff}}) \left[ 2 k_{\rm{B}} T_{\rm{n}} + \frac{1}{2}\mu_{\rm{n}} \left( {\bf{u}}_{{\rm{g}},Z} - {\bf{u}}_{\rm{n}} \right)^2 - 2 k_{\rm{B}} T_{\rm{d}} \right],
\label{grain_heat_neutral_coll}
\end{equation}
\noindent
where $\alpha_{\rm{d}}(T)$ is the accommodation coefficient, $T_{\rm{eff}}$ is calculated using the equation (\ref{eff_temp_gas_dust}). The accommodation coefficient $\alpha_{\rm{d}}(T)$ is taken to have a linear dependence on gas temperature logarithm, $\alpha_{\rm{d}}(T) = 0.35$ at $T = 10$~K and $\alpha_{\rm{d}}(T) = 0.1$ at $T = 10^4$~K \citep{Burke1983}. At higher gas temperatures, $\alpha(T)$ is set to be equal to 0.1. 

\subsection{Ion--neutral scattering}
Ions and electrons exchange momentum with neutral atoms and molecules via elastic scattering. For He and H scattering on ions, the 'classical' cross-section for momentum transfer in collisions between ions and neutral species is used \citep{Osterbrock1961}. For H$_2$ scattering on ions, the results of quantum mechanical calculations by \citet{Flower2000} are used. For the elastic scattering between electrons and neutral species (He, H, H$_2$), momentum transfer cross sections are adopted from \cite{Crompton1970,Dalgarno1999,Yoon2008}. The general expressions for the momentum transfer and heat source terms due to elastic scattering were given by \citet{Draine1986}. 

\subsection{Sputtering}
\label{sect_sputtering}
In C-type shocks, the gas remains relatively cold and sputtering is largely determined by the motion of dust grains through the neutral gas component \citep{Draine1995}. Sputtering rates depend on the sputtering yield $\langle Y(E) \rangle_{\rm{\theta,jk}}$ -- average number of sputtered particles $j$ per incident particle $k$, where $E$ is the energy of projectile particle, the brackets denote the averaging over the angle of incidence. We consider the sputtering of grain mantles by dominant neutral species H$_2$, He and CO. The sputtering yields are calculated according to \citet{Draine1979}.  

The sputtering rate of species $j$ is:

\begin{equation}
k_{\rm{sp, j}} = \frac{\langle Y u \rangle_{\rm{jk}} n_{\rm{k}}}{4 N_{\rm{act}} N_{\rm{s}}} \ \left[ {\rm{s}^{-1}} \right],
\end{equation} 

\noindent
where $n_{\rm{k}}$ is the projectile particle number density, $\langle Y u \rangle_{\rm{jk}}$ is the product of the sputtering yield and grain--projectile velocity difference averaged over the velocity distribution of projectile particles \citep{Draine1979}:

\begin{equation}
\begin{array}{l}
\displaystyle
\langle Y u \rangle_{\rm{jk}} = \left( \frac{8 k_{\rm{B}} T_{\rm{n}}}{\pi m_{\rm{k}}} \right)^{1/2} 
\frac{\rm{exp}(-s^2)}{s} \int\limits_0^{\infty} {\rm{d}}x \, \langle Y(x)\rangle_{\theta, {\rm{jk}}} \, x^2 \,{\rm{sinh}}(2xs)\,{\rm{exp}}(-x^2), \\ [15pt]
\displaystyle
x = \sqrt{\frac{E}{k_{\rm{B}} T_{\rm{n}}}}, \quad s^2 = \frac{m_{\rm{k}} \left({\bf{u}}_{\rm{g}} - {\bf{u}}_{\rm{n}} \right)^2}{2 k_{\rm{B}} T_{\rm{n}}},
\end{array}
\label{yield_velocity_product}
\end{equation}

\noindent
where $m_{\rm{k}}$ is the mass of the projectile particle. The integral in the equation (\ref{yield_velocity_product}) is calculated for a grid of parameters $s$ and $T_{\rm{n}}$ for each target--projectile pair. The destruction of molecules in the sputtering process and the sputtering of dust grain cores are not considered in our calculations.

\subsection{Numerical methods}
In shock simulations, the quantities to be integrated are: velocities of neutral and ion fluids $u_{\rm{n}}$, $u_{\rm{i}}$; kinetic temperatures of gas components $T_{\rm{n}}$, $T_{\rm{i}}$, $T_{\rm{e}}$; dust temperature $T_{\rm{d}}$; number densities of chemical species $n_{\rm{j}}$ and $n_{\rm{j}}^{\rm{s}}$; the grain charge distribution $n_{{\rm{g}},Z}$ (or the grain concentration $n_{\rm{g}}$ and average charge $\langle Z \rangle$); population densities of energy levels of ions and molecules $\chi_{\rm{jm}}$. The total number of coupled differential equations is approximately 1100. The solution starts from some initial values of flow variables except for ion fluid speed $u_{\rm{i}}$, which is given a small negative perturbation from its initial value, $u_{\rm{i}} = (1-\delta) u_{\rm{s}}$, where $\delta = 10^{-4}$ \citep{Flower1985}. The numerical integration is terminated when the relative difference between velocities of ion and neutral fluids is less than 0.1 per cent. We use differential equation solver CVODE v2.9.0, which is available in the web\footnote{\url{https://computation.llnl.gov/projects/sundials}} \citep{Hindmarsh2005}. The relative error of the computations is set equal to $10^{-6}$. It takes about 10--15~hours to calculate the shock model on 8 processor cores. The simulations were performed on the computer cluster of the Saint Petersburg branch of the Joint Supercomputer Centre of the Russian Academy of Sciences\footnote{\url{http://scc.ioffe.ru/}}.

The optical depths in atomic and molecular lines and, hence, rates of radiative cooling depend on the values of velocity gradient of neutral and ion fluids, which, in turn, depend on cooling rates \citep{Flower2012}. In calculations of optical depths, a some fixed value of velocity gradient is used. This value is reassigned when the relative difference between this parameter and actual velocity gradient becomes high ($> 10$ per cent).

\section{Collisional dissociation reactions}
\label{app:colldissreactions}
The chemical reaction network must be supplemented with collisional dissociation reactions that for the most part are absent in publicly available chemical networks. Table \ref{table:colldissreactions} presents a list of collisional dissociation reactions that are added to the chemical network. Usually the collisional partner in rate measurements of such reactions is Ar, N$_2$ or He. Keeping in mind that the most likely collisional partner in interstellar molecular gas is H$_2$, we crudely rescale collisional rates multiplying by the factor of 3 if collisional partner is Ar, N$_2$ or another heavy molecule, and do not rescale in the case of He \citep{Palau2017}. This factor is already taken into account in rate constants given in the Table \ref{table:colldissreactions}. The rate constants for collisional dissociation of CH, OH, H$_2$O and O$_2$ are already presented in UDfA chemical network.

\begin{table}
\small
\caption{Collisional dissociation reactions added to the chemical network.}
\begin{tabular}{llllll}
\hline
Reaction & $\alpha$ (cm$^{3}$ s$^{-1}$) & $\beta$ & $\gamma$ (K) & T(K) & Reference \\
\hline
O$_2$H + M $\to$ O$_2$ + H + M & 2.5$\times 10^{-8}$ & -0.76 & 24350 & 300--2000 & \citet{Tsang1986} \\
H$_2$O$_2$ + M $\to$ OH + OH + M & 9.7$\times 10^{-8}$ & 0 & 21990 & 930--1250 & \citet{Sajid2014} \\
CH$_2$ + M $\to$ CH + H + M & 2$\times 10^{-8}$ & 0 & 41800 & 2500--3800 & \citet{Dean1992} \\
CH$_2$ + M $\to$ C + H$_2$ + M & 6.5$\times 10^{-10}$ & 0 & 29700 & 2500--3800 & \citet{Dean1992} \\
CH$_3$ + M $\to$ CH + H$_2$ + M & 1.5$\times 10^{-8}$ & 0 & 40700 &  2710--3530 & \citet{Vasudevan2007} \\
CH$_3$ + M $\to$ CH$_2$ + H + M & 1.1$\times 10^{-8}$ & 0 & 41600 & 2250--2980 & \citet{Vasudevan2007} \\
CH$_4$ + M $\to$ CH$_3$ + H + M & 2.3$\times 10^{-6}$ & 0 & 45700 & 1000--1700 & \citet{Baulch2005} \\
 & 1.1$\times 10^4$ & -8.2 & 59200 & 1700--5000 & \citet{Baulch2005} \\ 
C$_2$ + M $\to$ C + C + M & 7.5$\times 10^{-8}$ & 0 & 71650 & 2580--4650 & \citet{Kruse1997} \\
C$_2$H + M $\to$ C$_2$ + H + M & 0.15 & -5.16 & 57367 & 2580--4650 & \citet{Kruse1997} \\
C$_2$H$_2$ + M $\to$ C$_2$H + H + M & 34 & -6.06 & 67130 & 2580--4650 & \citet{Kruse1997} \\
C$_2$H$_3$ + M $\to$ C$_2$H$_2$ + H + M & 1.4$\times 10^{-5}$ & -3.5 & 18070 & 200--2000 & \citet{Baulch2005} \\
C$_2$H$_4$ + M $\to$ C$_2$H$_2$ + H$_2$ + M & 3.1$\times 10^{-4}$ & 1 & 39390 & 1500--3200 & \citet{Baulch2005} \\
C$_2$H$_4$ + M $\to$ C$_2$H$_3$ + H + M & 1.3$\times 10^{-6}$ & 0 & 48600 & 1500--3200 & \citet{Baulch2005} \\
C$_2$H$_5$ + M $\to$ C$_2$H$_4$ + H + M & 5.1$\times 10^{-6}$ & 0 & 16800 & 700--900 & \citet{Baulch2005} \\
CH$_3$CH$_3$ + M $\to$ CH$_3$ + CH$_3$ + M & 1.4$\times 10^{5}$ & -8.37 & 47290 & 300--2000 & \citet{Baulch2005} \\
CN + M $\to$ C + N + M & 1.3$\times 10^{-9}$ & 0 & 71000 & 4060--6060 & \citet{Tsang1992} \\
HCN + M $\to$ CN + H + M & 6.5$\times 10^{-4}$ & -2.6 & 62850 & 2200--5000 & \citet{Tsang1991} \\
N$_2$ + M $\to$ N + N + M & 2.9$\times 10^{-4}$ & -3.33 & 113220 & 3390--6440 & \citet{Thielen1986} \\
NH + M $\to$ N + H + M & 9$\times 10^{-10}$ & 0 & 37650 & 2500--3400 & \citet{Deppe1998} \\
NH$_2$ + M $\to$ NH + H + M & 6$\times 10^{-9}$ & 0 & 38250 & 2200--4000 & \citet{Deppe1998} \\
NH$_3$ + M $\to$ NH$_2$ + H + M & 4.7$\times 10^{-8}$ & 0 & 46860 & 2000--3000 & \citet{Baulch2005} \\
NH$_3$ + M $\to$ NH + H$_2$ + M & 4.7$\times 10^{-8}$ & 0 & 46860 & 2000--3000 & \citet{Baulch2005} \\
NO + M $\to$ N + O + M & 4.8$\times 10^{-9}$ & 0 & 74700 & 2400--6200 & \citet{Tsang1991} \\
CO + M $\to$ C + O + M & 4.5$\times 10^{-4}$ & -3.1 & 129000 & 5500--9000 & \citet{Mick1993} \\
CO$_2$ + M $\to$ CO + O + M & 1.8$\times 10^{-9}$ & 0 & 52525 & 2400--4400 & \citet{Burmeister1990} \\
HCO + M $\to$ CO + H + M & 2$\times 10^{-10}$ & 0 & 7820 & 500--2500 & \citet{Baulch2005} \\
H$_2$CO + M $\to$ HCO + H + M & 2.4$\times 10^{-8}$ & 0 & 38050 & 1700--3000 & \citet{Baulch2005} \\
H$_2$CO + M $\to$ CO + H$_2$ + M & 1.4$\times 10^{-8}$ & 0 & 32100 & 1700--3000 & \citet{Baulch2005} \\
NCO + M $\to$ CO + N + M & 1.1$\times 10^{-9}$ & 0 & 27200 & 2000--3100 & \citet{Baulch2005} \\
HNCO + M $\to$ CO + NH + M & 4.9$\times 10^{-8}$ & 0 & 43000 & 1830--3340 & \citet{Mertens1989} \\
CH$_2$OH + M $\to$ H$_2$CO + H + M & 1.4$\times 10^{-3}$ & -5.39 & 18217 & 300--2500 & \citet{Baulch2005} \\
CH$_3$O + M $\to$ H$_2$CO + H + M & 3.5$\times 10^{-6}$ & -3 & 12230 & 500--1000 & \citet{Baulch2005} \\
CH$_3$OH + M $\to$ CH$_3$ + OH + M & 2.5$\times 10^{-7}$ & 0 & 33080 & 1000--2000 & \citet{Baulch2005} \\
CH$_3$OH + M $\to$ CH$_2$OH + H + M & 8.3$\times 10^{-8}$ & 0 & 33080 & 1000--2000 & \citet{Baulch2005} \\
CH$_3$OCH$_2$ + M $\to$ H$_2$CO + CH$_3$ + M & 1.4$\times 10^{-7}$ & 0 & 9110 & 473--573 & \citet{Loucks1967} \\
CH$_3$OCH$_3$ + M $\to$ CH$_3$O + CH$_3$ + M & 3.8$\times 10^{-8}$ & 0 & 21537 & 1350--1800 & \citet{Cook2009} \\
HC$_2$O + M $\to$ CO + CH + M & 3$\times 10^{-8}$ & 0 & 29600 & 1500--2500 & \citet{Frank1988} \\ 
CH$_2$CO + M $\to$ CO + CH$_2$ + M & 1.2$\times 10^{-8}$ & 0 & 28990 & 1650--1850 &  \citet{Frank1986} \\
CH$_3$CO + M $\to$ CO + CH$_3$ + M & $10^{-8}$ & 0 & 7080 & 400--500 & \citet{Baulch2005} \\
HOCO + M $\to$ CO + OH + M & 7.9$\times 10^{-5}$ & -2.4 & 18862 & 1400--2600 & \citet{Golden1998} \\
HOCO + M $\to$ CO$_2$ + H + M & 4.7$\times 10^{-5}$ & -3.15 & 18629 & 1400--2600 & \citet{Golden1998} \\
HCOOH + M $\to$ CO + H$_2$O + M & $10^{-9}$ & 0 & 20300 & 1370--2000 & \citet{Saito1984} \\
HCOOH + M $\to$ CO$_2$ + H$_2$ + M & 7.5$\times 10^{-8}$ & 0 & 28700 & 1280--2030 & \citet{Hsu1982} \\
CH$_3$CHO + M $\to$ CH$_3$ + HCO + M & 5$\times 10^{-4}$ & 0 & 37040 & 775--809 & \citet{Trenwith1963} \\
HCOOCH$_3$ + M $\to$ CH$_2$CO + H$_2$O + M & 1.4$\times 10^{-12}$ & 0 & 17200 & 733--868 & \citet{Blake1968} \\
C$_2$H$_5$OH + M $\to$ C$_2$H$_5$ + OH + M & 3$\times 10^{16}$ & -19.7 & 57600 & 800--1800 & \citet{Tsang2004} \\
C$_2$H$_5$OH + M $\to$ CH$_3$ + CH$_2$OH + M & 1.5$\times 10^{15}$ & -18.9 & 52750 & 800--1800 & \citet{Tsang2004} \\
C$_2$H$_5$OH + M $\to$ C$_2$H$_4$ + H$_2$O + M & 5.5$\times 10^{12}$ & -17.9 & 42650 & 800--1800 & \citet{Tsang2004} \\
CH$_3$CHCH$_2$ + M $\to$ CH$_3$ + C$_2$H$_3$ + M & 3.5$\times 10^{13}$ & -15.7 & 60400 & 1650--2300 & \citet{Kiefer1982} \\
\hline
\end{tabular}
\label{table:colldissreactions}
\end{table}

\end{document}